\documentclass[12pt,preprint]{aastex}
\newcommand{\kms}{{\,\rm km\,s}^{-1}}

\newcommand{\nodatl}{\multicolumn{1}{l}{~~~$\cdots$}}
\def\la{\mathrel{\hbox{\rlap{\hbox{\lower4pt\hbox{$\sim$}}}\hbox{$<$}}}}
\def\sun{\hbox{$\odot$}}
\renewcommand{\mag}{\mbox{$\;$mag}}

\begin{document}

\title{THE HUBBLE CONSTANT: A SUMMARY OF THE HST PROGRAM FOR THE 
             LUMINOSITY CALIBRATION OF TYPE Ia SUPERNOVAE 
                      BY MEANS OF CEPHEIDS} 
\author{A. Sandage}
\affil{The Observatories of the Carnegie Institution of Washington,\\
       813 Santa Barbara Street, Pasadena, CA 91101}
\author{G. A. Tammann}
\affil{Astronomisches Institut der Universit\"at Basel,\\
       Venusstrasse 7, CH-4102 Binningen, Switzerland}
\email{G-A.Tammann@unibas.ch}
\author{A. Saha}
\affil{NOAO, P.O. Box 26732, Tucson, AZ 85726}
\email{saha@noao.edu}
\author{B. Reindl}
\affil{Astronomisches Institut der Universit\"at Basel,\\
       Venusstrasse 7, CH-4102 Binningen, Switzerland}
\email{reindl@astro.unibas.ch}
\author{F. D. Macchetto, N. Panagia}
\affil{Space Telescope Science Institut, \\
       3700 San Martin Drive, Baltimore, MD 21218}
\email{macchetto@stsci.edu, panagia@stsci.edu}

\begin{abstract}
This is the fifth and final summary paper of our 15 year program using 
the Hubble Space Telescope (HST) to determine the Hubble constant using 
Type Ia supernovae, calibrated with Cepheid variables in nearby 
galaxies that hosted them. Several developments not contemplated at the 
start of the program in 1990 have made it necessary to put the summary 
on $H_{0}$ on a broader basis than originally thought, making four
preparatory papers (cited in the text) necessary.
   The new Cepheid distances of the subset of 10 galaxies, which were 
hosts of normal SNe\,Ia, give weighted mean luminosities in $B$, $V$, 
and $I$ at maximum light of $-19.49$, $-19.46$, and $-19.22$,
respectively. These calibrate the adopted SNe\,Ia Hubble diagram from
Paper~III to give $H_{0}=62.3\pm1.3$ (random) $\pm5.0$ (systematic) in
units of $\kms$\,Mpc$^{-1}$. This is a global value because it uses
the Hubble diagram between redshift limits of $3000$ and $20\,000\kms$
reduced to the CMB kinematic frame, well beyond the effects of any
local random and streaming motions.
   Local values of $H_{0}$ between 4.4 and 30$\;$Mpc from Cepheids,
SNe\,Ia, 21\,cm-line widths, and the tip of the red-giant branch
(TRGB) all agree within 5\% of our global value. 
   This agreement of $H_{0}$ on all scales from $\sim\!4 - 200$\,Mpc
finds its most obvious explanation in the smoothing effect of vacuum
energy on the otherwise lumpy gravitational field due to the
non-uniform distribution of the local galaxies. The physical methods
of time delay of gravitational lenses and the Sunyaev-Zeldovich effect
are consistent (but with large errors) with our global value. The
present result is also not in contradiction with existing analyses of
CMB data, because they either lead to wide error margins of $H_{0}$ or
depend on the choice of unwarrented priors that couple the value of
$H_{0}$ with a number of otherwise free parameters in the CMB acoustic
waves. Our value of $H_{0}$ is 14\% smaller than the value of $H_{0}$
found by \citet{Freedman:etal:01} because our independent Cepheid
distances to the six SNe\,Ia-calibrating galaxies used in that
analysis average $0.35\mag$ larger than those used earlier.
\end{abstract}
\keywords{Cepheids --- distance scale --- galaxies: distances and
  redshifts --- supernovae: general}

\section{INTRODUCTION}
\label{sec:intro}
Supernovae of Type Ia (SNe\,Ia) are uniquely suited for the
calibration of the Hubble constant $H_0$ because they are accessible
out to large distances, and as the best standard candles known they
are insensitive to selection (Malmquist) bias, which has beset the
extragalactic distance scale for so long. An HST program to determine
the Hubble constant $H_0$ using Type Ia supernovae (SNe\,Ia) as
standard candles was therefore mounted in 1990 at the time when HST
was launched. 
The plan was to determine Cepheid distances of nearby galaxies which
had produced well observed SNe\,Ia, and to compare their resulting
mean absolute magnitudes with the apparent magnitudes of SNe\,Ia out
to $\sim\!30\,000\kms$ (in the present context referred to as
``distant'' SNe\,Ia). This velocity distance is ideal for the
determination of the-large-scale value of $H_0$(cosmic) because it is
large enough that any noise in the Hubble flow due to random
velocities and local streaming motions of galaxies is negligible
compared with the systematic redshift itself, and on
the other hand is small enough that cosmological effects are also
negligible (except for small $K$-corrections, see below).

     The program progressed step by step. By 2001 we had 
published with our collaborators the Cepheid distances 
of eight SNe\,Ia-bearing galaxies 
\citep*[for references see][hereafter Paper~IV]{Saha:etal:06}. 
Four additional Cepheid distances of such galaxies are due to other
authors. Thus Cepheid distances are now available for 12 SNe\,Ia of
which, however, two are spectroscopically peculiar.

     Yet at least four originally unforeseen developments have
made it necessary to put the present summary paper on $H_0$ on a
broader basis, which we have prepared in four preceding papers
(\citealt*{Tammann:etal:03}, hereafter Paper~I; 
 \citealt*{Sandage:etal:04}, hereafter Paper~II; 
 \citealt*{Reindl:etal:05},  hereafter Paper~III; and
 \citealt*{Saha:etal:06},    hereafter Paper IV). 
The four developments are as follows:

     (1) The uniformity of SNe\,Ia. While some pushed SNe\,I 
\citep{Kowal:68,Barbon:etal:75,Branch:77,Branch:Bettis:78,Tammann:79,Tammann:82},
or particularly SNe\,Ia 
\citep{Cadonau:etal:85,Leibundgut:88,Leibundgut:91,Sandage:Tammann:90,Branch:Tammann:92}
as standard candles with the then available 
observations, others emphasized their variety
\citep{Pskovskii:67,Pskovskii:84,Barbon:etal:73,Frogel:etal:87}. 
Eventually \citet{Phillips:93}
proposed a correlation between the decline rate $\Delta m_{15}$ (the
magnitude change during  the first 15 days past $B$ maximum) and the
absolute magnitude at maximum. The quantitative correlation became
convincing when \citet{Hamuy:etal:96} showed that the magnitude
residuals from the Hubble diagram regression (Hubble line) are a
function of $\Delta m_{15}$.   
There remained some debate between different authors as to the
steepness of the correlation. But this is now understood as the result
of different assumptions on the intrinsic color of SNe\,Ia; the
$\Delta m_{15}$ correction on the magnitudes is therefore quite well
controlled (\citeauthor{Reindl:etal:05}), but remains purely
empirical. It became also increasingly clear that spectroscopically
peculiar SNe\,Ia cannot be used as standard candles.
The magnitude of normal SNe\,Ia after correction for 
$\Delta m_{15}$ show a scatter about the Hubble line of 
$\sigma=0.15\mag$ (including errors of the absorption corrections; 
the {\em intrinsic\/} scatter is demonstrably $\le0.11\mag$). 
Without the $\Delta m_{15}$ correction a scatter of $0.21\mag$ would
be observed which shows -- contrary to occasional claims -- that even
without the $\Delta m_{15}$ correction SNe\,Ia are highly competitive
standard candles. 
The $\Delta m_{15}$ corrections do not only affect the
observed magnitude scatter, but also the calibration of $H_0$ at the
level of 5\% because the calibrating SNe\,Ia have systematically
smaller $\Delta m_{15}$ values than the distant SNeIa. 
[This is because Cepheids require galaxies with a young population,
and the SNe\,Ia in such galaxies, i.e.\ in spirals as contrasted to
the ellipticals, tend to have small $\Delta m_{15}$ as first pointed
out by \citet{Hamuy:etal:95}]. -- It is 
self-understood that the calibration of $H_0$  requires a large
sample of uniformly reduced distant SNe\,Ia; this was compiled in
\citeauthor{Reindl:etal:05}. The large size of the sample is due to
the heroic effort of many observers who have collected the necessary
photometry over the years.  

     (2) The difference of the $P$-$L$ relation of Cepheids in the Galaxy
and in LMC. In Paper~I \& II it was shown that not only do the Galactic
period-color ($P$-$C$) relations in $(B\!-\!V)$ and $(V\!-\!I)$ have
different slopes than in the LMC, but -- as a necessary
consequence -- the period-luminosity ($P$-$L$) relations 
in $B$, $V$, and $I$ have different slopes, and hence different
luminosities at given periods, in the two galaxies.
Confirmation, among other indicators, is that the lightcurve shape at
a given period differ between Cepheids in the Galaxy and the LMC
\citep{Tanvir:etal:05}. 

This came as a rather unexpected complication for the determination of
Cepheid distances. It requires that the $P$-$L$ relations of the two
galaxies must be based on independent zero-points. An update of the
zero-points adopted in Paper~II and IV and their errors is given in
\S~\ref{sec:H0cosmic:error}~(2b).
The inequality of the $P$-$L$ relations raises a deep problem:
which relation, or an interpolation between them, is to be used for a
particular galaxy? 

     (3) In \citeauthor{Saha:etal:06} it was decided to determine 
{\em two\/} Cepheid distances for each SN\,Ia host galaxies, once 
$\mu^{0}$(Gal) with the $P$-$L$ relation of the relatively metal-rich
Galactic Cepheids, and once $\mu^{0}$(LMC) with the $P$-$L$ relation of
the metal-poor LMC Cepheids, and then to interpolate -- and slightly
extrapolate -- between the two values according to metallicity,
expressed by the new [O/H] scale of \citet{Sakai:etal:04}, 
hereafter denoted by $[\mbox{O/H}]_{\rm Sakai}$. 
The resulting moduli $ \mu^{0}_{Z} $ are in this way corrected for
metallicity, provided the full difference between the Galactic and LMC
$P$-$L$ relation is in fact caused by metallicity differences. 
In \citeauthor{Sandage:etal:04} only part of the difference between
$\mu^{0}$(Gal) and $\mu^{0}$(LMC) could be proved to be caused by the
line blanketing effect of metal lines, 
but most of the difference could not be explained. Thus it remained a
hypothesis that {\em all\/} of the difference is due to metallicity
variations. 

     Perhaps the first to predict that the position of the edges of
the Cepheid instability strip is a function of chemical composition
(in its effect on the atmospheric opacity) was  John Cox
(\citeyear{Cox:59}; \citeyear{Cox:80}, eq.~10.4). 
The sense was that the strip boundaries move
bluewards with decreasing $Z$, yet redwards with decreasing $Y$. 
Hence, for a particular $\Delta Y/ \Delta Z$, if $Y$ increases with
increasing $Z$, there will be compensation in the $L-T_{\rm e}$
instability line, and the strip boundary is in this case almost
independent of variable $Y$ and $Z$ for that value of 
$\Delta Y/ \Delta Z$. 
Among the first to calculate the effect were \citet{Christy:66},
\citet{vanAlbada:Baker:71}, and \citet{Tuggle:Iben:72}.
\citet{Sandage:etal:99} used the modern models of
\citet{Chiosi:etal:92,Chiosi:etal:93} to show that 
the effects of $Y$ and $Z$ nearly compensate when
$\Delta Y/ \Delta Z \sim 5.5$ (see eq.~50 of
\citeauthor{Sandage:etal:04}).

     The situation has become much clearer due to recent model
calculations by \citet{Fiorentino:etal:02} and \citet{Marconi:etal:05}
who show that the strip moves bluewards with decreasing $Z$,
irrespective of any reasonable value of $Y$.
The models of \citet{Marconi:etal:05} can match the observed $P$-$L$
relation of LMC, {\em including its break at $\log P=1$}, impressively
well, except that the model colors $(V\!-\!I)$ are still too red by
$\sim\!0.1\mag$ at the long periods ($\langle\log P\rangle\approx1.5$)
needed for Cepheids outside the Local Group. A corresponding color
difference persists for $\langle\log P\rangle=1.5$ also between the
observed and theoretical Galactic $P$-$L$ relations in $V$ and
$I$. Yet the main point here is that the shift of the instability
strip to higher temperatures with decreasing metallicity is no longer
a hypothesis, but is predicted by model calculations.

     Empirical support for the metallicity corrections comes from
comparing the resulting $\mu^{0}_{Z}$ of 37 galaxies in
\citeauthor{Saha:etal:06} with independent TRGB distances, as far as
available, and with velocity distances. The distances
$\mu^{0}_{Z}-\mu^{0}_{\rm TRGB}$ and $\mu^{0}_{Z}-\mu^{0}_{\rm vel}$
show no significant dependence on [O/H]. A clear dependence does
arise, however, if $\mu^{0}$(Gal) or $\mu^{0}$(LMC) were used instead
of $\mu^{0}_{Z}$. Also the strong metal dependence of the SN\,Ia
luminosities based on $\mu^{0}$(LMC) becomes insignificant if 
$\mu^{0}_{Z}$ is used instead (\citeauthor{Saha:etal:06}).

     The ten nearby galaxies with Cepheid distances that have hosted
normal SNe\,Ia carry metallicity corrections, which were
found (\citeauthor{Saha:etal:06}) to depend not only on the
metallicity, but also on the mean period of the Cepheids. 
The corrections vary for $\mu^{0}$(Gal) between $-0.30$
and $+0.10\mag$, and for $\mu^{0}$(LMC) between $-0.11$ and
$+0.36\mag$. The {\em mean\/} corrections are
$\langle\Delta\mu_{Z}\rangle =-0.02$ for $\mu^{0}$(Gal) and
$+0.20\mag$ for $\mu^{0}$(LMC). The small mean correction in case of
$\mu^{0}$(Gal) is due to the fact that the mean metallicity of
[O/H]$_{\rm Sakai}=8.55$ of the ten calibrating galaxies is
close to the metallicity of the Galactic Cepheids ($8.60$). 
(The notation [O/H]$_{\rm Sakai}$ is explained in \S~\ref{sec:data}).
Therefore, if one is prepared to accept the simplifying premise that
the distances of metal-rich Cepheids should be based on the $P$-$L$
relation of the metal-rich Galactic Cepheids -- without any further
metallicity correction -- one obtains on average a good approximation
(to within $0.02\mag$ or 1\% in distance) to the adopted $\mu^{0}_{Z}$
of the calibrators, and hence to the mean luminosity of their SNe\,Ia. 

     (4) A new photometric zero-point of the HST WFPC2 camera was
determined in \citeauthor{Saha:etal:06}. It affects six of the present
calibrators by $0.02$ to $0.07\mag$ depending on the chip and the
epoch of observation. The photometric zero-point of the remaining four
galaxies was estimated to deviate by not more than $0.05\mag$.

The structure of the paper is as follows:
     The basic data of twelve SNe\,Ia (of which ten are normal
SNe\,Ia) and their host galaxies are compiled from Papers~III and IV
in \S~\ref{sec:data}. 
The mean absolute magnitudes $M_{BVI}$ at maximum of the normal
SNe\,Ia are derived in \S~\ref{sec:Mabs}.  
The values $M_{BVI}$ are combined with the Hubble diagram of more
distant SNe\,Ia to yield in \S~\ref{sec:H0cosmic} the large-scale
value of $H_{0}$ and its error. 
The local value of $H_{0}$ within $2000\kms$ is derived from Cepheids,
SNe\,Ia, 21cm-line width and tip of the red-giant branch (TRGB)
distances in \S~\ref{sec:H0local}.   
The evidence for $H_{0}$ from physical distance determinations is
briefly discussed in \S~\ref{sec:physical}.
In \S~\ref{sec:conclusions} the conclusions are given.

\section{THE BASIC DATA OF THE CALIBRATING SNe\,Ia AND OF THEIR HOST GALAXIES}
\label{sec:data}
The relevant parameters of the 12 galaxies with SNe\,Ia and known
Cepheid distances are compiled in Table~\ref{tab:01} from
\citeauthor{Saha:etal:06} (Table~A1). Columns~(3) and (4) of
Table~\ref{tab:01} list their recession velocities, corrected
to the barycenter of the Local Group
\citep{Yahil:etal:77} and for a self-consistent Virgocentric infall
model with a local infall vector of $220\kms$
\citep{Yahil:etal:80,Tammann:Sandage:85,Kraan-Korteweg:86}. 
The metallicities [O/H]$_{\rm old}$
from \citet{Kennicutt:etal:98} and others, as compiled by
\citet{Ferrarese:etal:00} are in column~(5).  The metallicities
[O/H]$_{\rm Sakai}$ in column~(6) are the $T_{\rm e}$-based
values introduced by \citet{Sakai:etal:04} and used in
\citeauthor{Saha:etal:06}. In 
this system Cepheids in the Galaxy have $[\mbox{O/H}]=8.6$
\citep{Andrievsky:etal:02} on average, and those in LMC 8.34
\citep{Sakai:etal:04}. Column~(7) gives the number of Cepheids which
enter the various distance determinations. 
The mean period of the accepted Cepheids is in column~(8).
Columns~(9)-(13) give the distance moduli derived in 
\citeauthor{Saha:etal:06} from different $P$-$L$ relations as follows:

(1)
The distance moduli $\mu^{0}$(Gal) in column~(9) are derived
from the {\em Galactic\/} $P$-$L$ relation (derived in
\citeauthor{Tammann:etal:03} and slightly revised in
\citeauthor{Sandage:etal:04}) {\em without\/} any metallicity
correction; their zero-point rests on the Pleiades at $\mu^{0}=5.61$
and with equal weight on moving-atmosphere parallaxes [the
Baade-Becker-Wesselink (BBW) method].

(2)
The distance moduli $\mu^{0}$(LMC) in column~(10) are derived
from the LMC $P$-$L$ relations (derived in \citeauthor{Sandage:etal:04})
again {\em without\/} any metallicity correction; their zero-point
rests on LMC at $\mu^{0}=18.54$, as justified in
\S~\ref{sec:H0cosmic:error}, but {\em excluding\/} all solutions based
on any $P$-$L$ relations. 

(3)
The distance moduli $\mu^{0}_{Z}$(M/F) in column~(11) are based
on the slopes of the old $P$-$L$ relations of \citet{Madore:Freedman:91},
adjusted to a zero-point at $\mu^{0}_{\rm LMC}=18.54$, and with the {\em
  period-independent\/} metallicity corrections from
\citeauthor{Saha:etal:06} (Table~7, col.~[6]); they correspond to
$\Delta\mu_{Z}=-0.65\Delta[\mbox{O/H}]_{\rm Sakai}$. 
[It was explained in \citeauthor{Saha:etal:06} why these corrections
are larger for the {\em long-period\/} Cepheids under
consideration than in \citet{Sakai:etal:04} who give 
$\Delta\mu_{Z}=-0.24\Delta[\mbox{O/H}]_{\rm old}$ or
$-0.32\Delta[\mbox{O/H}]_{\rm Sakai}$ based on Cepheids with all
periods].

(4)
The adopted distance moduli $\mu^{0}_{Z}$ in column~(12) (and
their errors in col.~[13]) are taken from
\citeauthor{Saha:etal:06}. They are a compromise between items (1) and
(2). The method is based on the assumption -- now confirmed by
\citet{Fiorentino:etal:02} and \citet{Marconi:etal:05}, as mentioned
before -- that their modulus
difference is a function of the metallicity {\em and\/} of the mean
period as expressed in equation~(10) in \citeauthor{Saha:etal:06}.

     In Table~\ref{tab:02} the relevant parameters of the 12 SNe\,Ia
of the sample are compiled from \citeauthor{Reindl:etal:05}. Column~(2)
lists the decline rates $\Delta m_{15}$, column~(3) the Galactic color
excess $E(B\!-\!V)_{\rm Gal}$ from \citet{Schlegel:etal:98}, and
column~(4) the color excess $E(B\!-\!V)_{\rm host}$ in the host
galaxy. For the Galactic absorption $A_{BVI}$ conventional
absorption-to-reddening ratios $4.1$, $3.1$, and $1.8$, respectively,
were assumed, while the corresponding values of $3.65$, $2.65$, and
$1.35$ were derived in \citeauthor{Reindl:etal:05} for the absorption
of the SNe\,Ia in the host galaxies. Columns~(5) and (6) give the
dereddened pseudo-colors $(B_{\rm max}-V_{\rm max})$ and 
$(V_{\rm max}-I_{\rm max})$.\footnote{For brevity we write in this  
  paper $(B\!-\!V)\equiv (B_{\rm max}-V_{\rm max})$ and  
  $(V\!-\!I)\equiv (V_{\rm max}-I_{\rm max})$. The values $(B\!-\!V)$   
  and $(V\!-\!I)$ are corrected for Galactic and internal  
  reddening. The designations $(B\!-\!V)^{\rm corr}$ and  
  $(V\!-\!I)^{\rm corr}$ mean that the colors are normalized in  
  addition to a decline rate of $\Delta m_{15}=1.1$.}  
The apparent magnitudes $m^{\rm corr}_{BVI}$ (and their adopted errors
in parentheses in units of $0.01\mag$; formal errors of $<0.05\mag$
are unrealistic because of the correction for absorption in the host
galaxy; they are set to $0.05\mag$.) in columns~(7)-(9) are corrected for
Galactic and host absorption and reduced to a common decline rate of
$\Delta m_{15}=1.1$ and to a common color of
$(B\!-\!V)_{\rm max}=-0.024$ [equation~(23) in
\citeauthor{Reindl:etal:05}]. 

     The apparent magnitudes $m^{\rm corr}_{BVI}$ of the calibrating
SNe\,Ia in Table~\ref{tab:02} are combined with the various distance
moduli in Table~\ref{tab:01} to yield the absolute magnitudes
$M_{BVI}$ in Table~\ref{tab:03}. Also shown in each column are the
straight and weighted mean values of $M_{BVI}$ for the ten normal
SNe\,Ia. The two spectroscopically peculiar type Ia supernovae
SN\,1991T, which is the prototype of an overluminous class, and 1999by
which belongs to the underluminous class of SN\,1991bg
(\citeauthor{Reindl:etal:05}) are listed separately; 
they are not further used in this paper.

\section{THE ABSOLUTE MAGNITUDE OF SNe\,Ia}
\label{sec:Mabs}
%
\subsection{The Range of Absolute Magnitudes}
\label{sec:Mabs:range}
The mean absolute magnitudes $M_{BV}$ of the ten calibrating SNe\,Ia
in Table~\ref{tab:03} vary between $-19.30$ and $-19.55$, the $M_{I}$
values vary between $-18.99$ and $-19.22$. These ranges translate into
a variation of $H_{0}$ of $10-15\%$. A careful scrutiny of the best
values of $M_{BVI}$ is therefore still necessary.

\subsection{The Adopted Absolute Magnitudes}
\label{sec:Mabs:adopted}
The faintest magnitudes in Table~\ref{tab:03} come from the distances
$\mu^{0}$(LMC) which are based on the LMC $P$-$L$ relations from
\citeauthor{Sandage:etal:04}. They are {\em uncorrected\/} for
metallicity as stated before. This is quite unrealistic because the
mean metallicity of the calibrating Cepheids
($\langle[\mbox{O/H}]_{\rm old}\rangle=8.79$, 
 $\langle[\mbox{O/H}]_{\rm Sakai}\rangle=8.55$) 
is significantly higher than that of LMC
($[\mbox{O/H}]_{\rm old}=8.50$, 
 $[\mbox{O/H}]_{\rm Sakai}=8.34$).
If one applies conservatively the period-independent metallicity
correction of \citet{Sakai:etal:04}, i.e.\ 
$\Delta\mu_{Z}=-0.24([\mbox{O/H}]_{\rm old}-8.50)$, the
metallicity-corrected $\mu^{0}_{Z}$(LMC) become larger by $0.07\mag$
{\em on average}. The resulting {\em weighted\/} absolute SN
magnitudes $M_{BVI}$ are shown in Table~\ref{tab:04}, line 2.
The corresponding average metallicity correction to the
$\mu^{0}$(Gal), and hence to the $M_{BVI}$, from the Galactic $P$-$L$
relation remains at the level of $\sim\!0.02\mag$ (to become fainter
in $B$ and $V$), because the mean metallicity of the calibrators is
almost as high as that of the Galaxy.

     The weighted absolute magnitudes $M_{BVI}$ for the four different
solutions in Table~\ref{tab:03} take now the values shown in
Table~\ref{tab:04}. 
The maximum difference within each column of Table~\ref{tab:04}
is $0.15\mag$. However, it must be noted that solution (2) has
the lowest weight, because it combines the $P$-$L$ relations of the
metal-poor LMC with the calibrating galaxies whose average metallicity
is much closer to the Galactic value. Eliminating therefore solution (2)
reduces the difference in each column to an almost negligible value of
$\le0.03\mag$. One could therefore argue for the mean magnitudes of
the three solutions. Yet for obvious reasons discussed in great detail
in \citeauthor{Saha:etal:06} we adopt the weighted magnitudes
$M_{BVI}$ of solution (4), where the period-dependent metallicity
corrections are applied. 
(It may seem surprising that the adopted magnitudes in
Table~\ref{tab:04} are almost as bright or in the case of $I$ even
brighter than those from the Galactic $P$-$L$ relation, although the
Galactic Cepheids are slightly more metal-rich than the mean value of
the calibrating galaxies ($\Delta [\mbox{O/H}]_{\rm Sakai}=0.05$) and
should give  -- with no metallicity correction applied -- somewhat
brighter SN\,Ia magnitudes. 
However, the exact differences in $M_{BVI}$ in Table~\ref{tab:04} are
modified by the weighting of individual SNe\,Ia. Moreover, 
the only six calibrating SNe\,Ia with $I$ magnitudes happen to lie in
relatively metal-rich galaxies).

\subsection{Tests of the Adopted Absolute Magnitudes}
\label{sec:Mabs:tests}
\begin{figure}[p] 
   \epsscale{0.65}
   \plotone{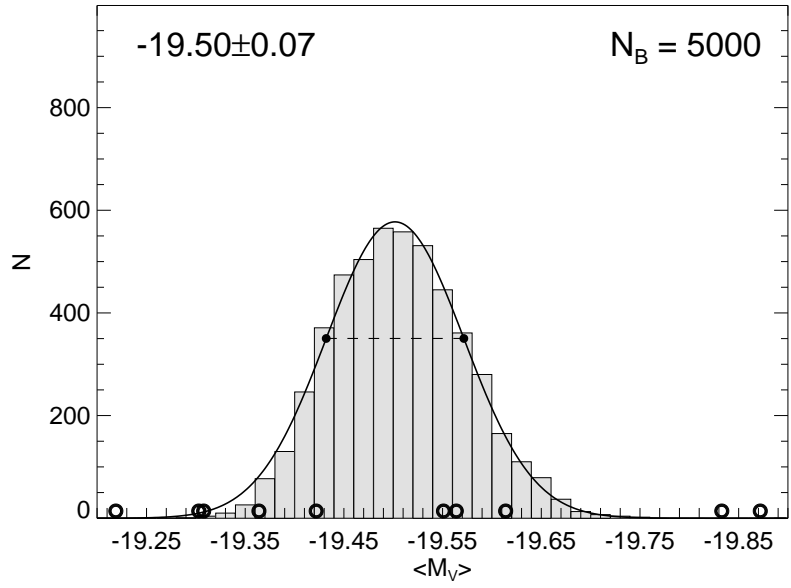}
   \caption{A bootstrap analysis of the adopted absolute SN magnitudes
   $M_{V}$. The individual SNe\,Ia are shown as open circles.} 
\label{fig:01}
\end{figure}

In order to test whether the 10 calibrating SNe\,Ia form a random
statistical sample a bootstrap analysis is applied to their $M_{V}$
magnitudes from Table~\ref{tab:03}, column~(12). 
A sample of 10 randomly chosen SNe\,Ia is formed from the list of 10,
allowing each SN to occur as many times as it is drawn, and the mean
value $\langle M_{V}\rangle$ is determined. This process is repeated
$m$ times; we have chosen $m=5000$. 
The distribution of the resulting mean magnitudes is shown in
Figure~\ref{fig:01}. The very good Gaussian fit to the observed
distribution argues for the SNe\,Ia magnitudes being randomly
distributed about a mean value. The mean value $M_{V}=-19.50$ is
identical to the unweighted mean in Table~\ref{tab:03}, column~(12).

     It is not meaningful to compare the present calibration of
$M_{V}=-19.50$ (unweighted) or $-19.46$ (weighted) with previous
authors, because different authors have used different precepts
concerning intrinsic color, color excess, and absorption-to-reddening
ratio ${\cal R}$; some of the published values are also normalized to
different values of the decline rate $\Delta m_{15}$ and the intrinsic
color $(B\!-\!V)^{0}$. Since these precepts have little effect on
$H_{0}$ as long as they are consistently applied to the nearby
calibrating SNe\,Ia and those defining the Hubble diagram, it is more
realistic to compare the values of $H_{0}$ from different authors.
This is done in \S~\ref{sec:H0cosmic:previous}.

\begin{figure*}[p] 
     \epsscale{0.8}
     \plotone{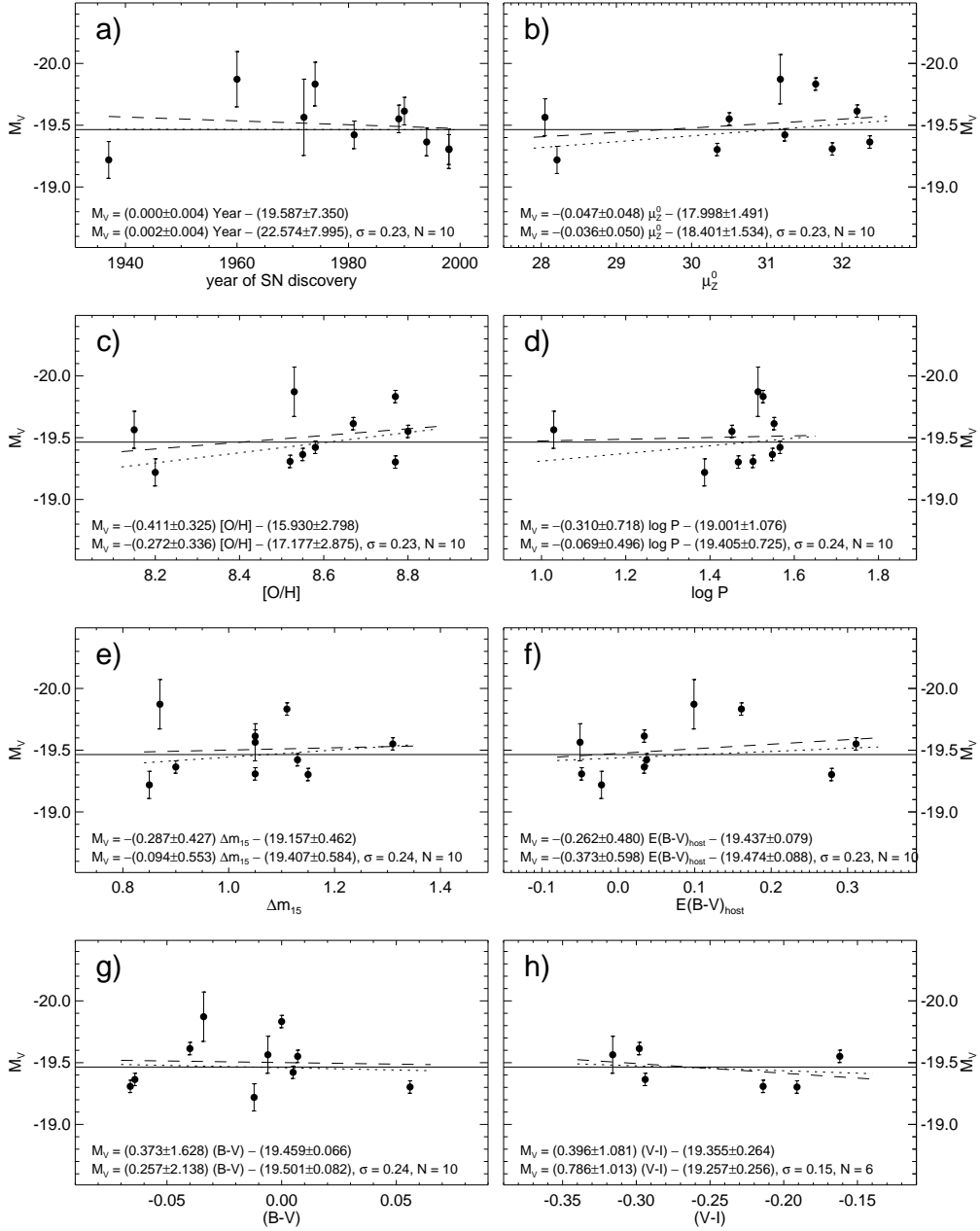}
     \caption{The correlation of the absolute SN magnitudes $M_{V}$ on
     a) year of discovery, b) distance modulus $\mu^{0}_{Z}$, c)
     metallicity [O/H]$_{\rm Sakai}$, d) mean period of the
     Cepheids $\langle\log P\rangle$, e) the decline rate $\Delta
     m_{15}$, f) the color excess in the host galaxy, g) the intrinsic
     color $(B\!-\!V)$, and h) the intrinsic color $(V\!-\!I)$.
     The upper equation in each panel gives the weighted regression
     (dotted line), the lower equation the unweighted regression
     (dashed line).}
\label{fig:02}
\end{figure*}

     To further ensure the homogeneity of the SN sample, their
magnitudes $M_{V}$ are plotted against various parameters in
Figure~\ref{fig:02}. This is to ascertain that the individual
magnitudes do or do not depend on age of the photometry, distance,
metallicity, mean period of the Cepheids, or decline rate $\Delta
m_{15}$, nor on the derived parameters like color excess or intrinsic
color. As can be seen in Figure~\ref{fig:02} there are no significant
trends, with the only exception of the weighted solution of $M_{V}$
versus [O/H]$_{\rm Sakai}$, which suggests a correlation at only
the $1.3\sigma$ level. We take the magnitude difference of
$0.08\pm0.15\mag$ (unweighted) or $0.13\pm0.09\mag$ (weighted)
between the five SNe\,Ia in metal-poor galaxies and their counterparts
in metal-rich galaxies as insignificant, because it depends entirely
on the assigned weights. It would therefore be arbitrary to exclude
one or more SNe\,Ia from the sample on the basis of any of the
parameters considered.
\clearpage
\clearpage

\clearpage
\clearpage

\section{THE LARGE-SCALE VALUE OF \boldmath{$H_{0}$}}
\label{sec:H0cosmic}
The value of $H_{0}$(cosmic) is obtained by combining the intercept
$C_{\lambda}$ of the Hubble line defined by distant SNe\,Ia with the
mean absolute magnitude of the ten calibrating SNe\,Ia. This is
because the three parameters are connected by (see
\citeauthor{Reindl:etal:05}): 
\begin{equation}
\log H_{0} = 0.2 M^{\rm corr}_{\lambda} + C_{\lambda} + 5,
\label{eq:01}
\end{equation}
where the magnitudes are corrected for the Galactic and internal
absorption, reduced to a common decline rate $\Delta m_{15}=1.1$ and
to a common color $(B-V)=-0.024$ at $\Delta m_{15}=1.1$.

\subsection{The Adopted Value of \boldmath{$H_{0}$}(cosmic)}
\label{sec:H0cosmic:adopted}
\begin{figure}[p] 
     \epsscale{0.65}
     \plotone{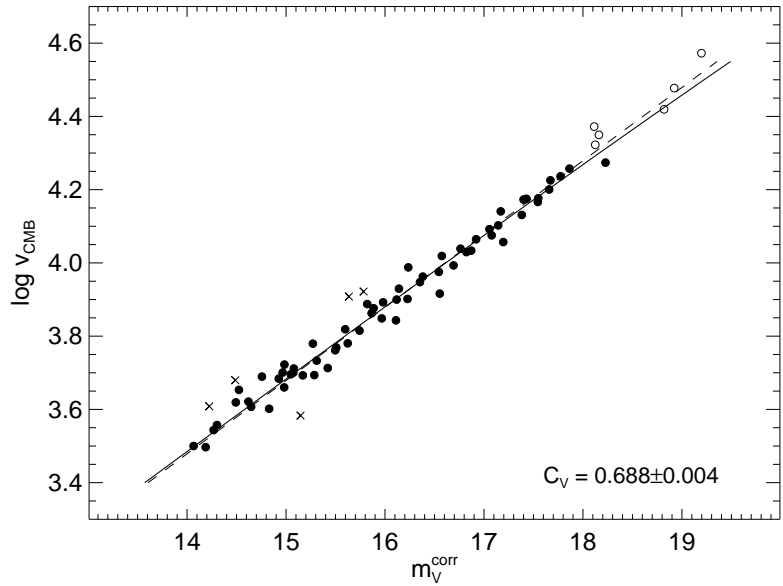}
     \caption{The Hubble diagram of 62 normal SNe\,Ia with
     $3000<v_{\rm cmb}\la30\,000\kms$. SNe\,Ia with $v_{\rm
     cmb}>20\,000\kms$ are shown as open symbols. Five possibly
     non-normal SNe\,Ia are shown as crosses. The dispersion is
     $\sigma=0.14\mag$. The dashed line holds for a $\Omega_{\rm M}=1$
     model. The data are from \citeauthor{Reindl:etal:05} of
     \citet*{Reindl:etal:05}.}
\label{fig:03}
\end{figure}

The Hubble diagram for a $\Omega_{\rm M}=0.3$, $\Omega_{\Lambda}=0.7$
model, using the apparent magnitudes $m^{\rm corr}_{V}$ of 62 normal
SNe\,Ia with $3000<v_{\rm cmb}<20\,000\kms$, is shown in
Figure~\ref{fig:03} from the data in
\citeauthor{Reindl:etal:05}. SNe\,Ia with $v<3000\kms$ are not 
considered because of the possible effect of local random and
streaming motions. Also the six SNe\,Ia with $20\,000<v_{\rm
  cmb}\la30\,000\kms$ are not used for the solution in order to avoid
large $K$-corrections; if they had been included they would decrease
$H_{0}$ by less than 1\%. Finally five possibly non-normal SNe\,Ia,
discussed in \citeauthor{Reindl:etal:05}, are excluded from the
solution.They are shown as crosses in Figure~\ref{fig:03}.

With these precepts one obtains for $C_{B}$, and correspondingly for
$C_{V}$ and $C_{I}$:
\begin{equation}
C_{B}=0.693\, (N=62),\; C_{V}=0.688\,(N=62),\; C_{I}=0.637\, (N=58).
\label{eq:02}
\end{equation}
The random error of the mean is in all three cases as small as
$0.004$. Inserting these values together with the adopted magnitudes
$M_{BVI}$ in Table~\ref{tab:04} yields
\begin{equation}
H_{0}(B)=62.4\pm1.2, \; H_{0}(V)=62.5\pm1.2, \;
H_{0}(I)=62.1\pm1.4,\footnote{\mbox{The units of $H_{0}$ are
    $\kms\;$Mpc$^{-1}$ throughout}}   
\label{eq:03}
\end{equation}
from which we adopt
\begin{equation}
H_{0}({\rm cosmic})=62.3\pm1.3.  
\label{eq:04}
\end{equation}
\clearpage
\clearpage
An estimate of the systematic error follows in \S~\ref{sec:H0cosmic:error}.
The close agreement of $H_{0}$ in all three colors speaks in favor of
the {\em consistent\/} absorption corrections applied to the
calibrators and distant SNe\,Ia in \citeauthor{Reindl:etal:05}. 
This is also seen in Table~\ref{tab:05}, where the mean intrinsic
colors (after correction for Galactic and internal reddening) of the
two sets of SNe\,Ia are compared. Their agreement is as good as can be
expected.  

     The color agreement between calibrators and distant SNe\,Ia is of
greatest importance. If it had not been the case it would have implied
that one set of SNe\,Ia had systematically larger absorption
corrections than the other. A biased value of $H_{0}$ would be the
result. Other intrinsic colors and hence absorption corrections
have been adopted by previous authors. This is unconsequential for
$H_{0}$ as long as the absorption corrections are consistent. But it
precludes a comparison of the present values of $C_{\lambda}$ with
previous authors. 

\subsection{\boldmath{$H_{0}$}(cosmic) from a Restricted Sample of SNe\,Ia}
\label{sec:H0cosmic:restricted}
As mentioned before the mean decline rate $\Delta m_{15}$ of the
calibrators is for good reasons smaller than that of the distant
SNe\,Ia (see also Table~\ref{tab:05}). The value of $H_{0}$ depends
therefore somewhat on the adopted $\Delta m_{15}$
corrections. Although the former disagreement between the corrections
by different authors are now essentially understood
(\citeauthor{Reindl:etal:05}), it is interesting to see the effect on
$H_{0}$ if the distant SNe\,Ia with $\Delta m_{15}> 1.28$ are
omitted. In that case the remaining 40 distant SNe\,Ia have exactly
the same $\langle\Delta m_{15}\rangle$ as the calibrators. They yield 
$C_{V}=0.687\pm0.005$ and hence $H_{0}(V)=62.4\pm1.2$.

     An alternative sample restriction is to dispense with the 
$\Delta m_{15}$ corrections altogether and to require -- since the
SN\,Ia luminosity is also a function of the type $T$ of the parent galaxy
(as defined in \citeauthor{Reindl:etal:05}; see Fig.~10 there) 
-- the mean type $T$ of the
parent galaxies be the same for the calibrators and distant
galaxies. Omitting the early-type galaxies with $T\le1$ leaves a
sample of 29 SNe\,Ia with $\langle T\rangle=3.9$. Their Hubble diagram
gives an intercept of $C_{V}=0.694\pm0.008$, from which follows
$H_{0}(V)=63.4\pm1.3$. The statistical error is here somewhat larger,
because the SN\,Ia luminosity does not correlate as tightly with $T$
as with $\Delta m_{15}$, but the essential point is that even without
the $\Delta m_{15}$ corrections $H_{0}$ does not change
significantly.

\subsection{Systematic Errors of \boldmath{$H_{0}$}}
\label{sec:H0cosmic:error}
The very small statistical error of $H_{0}$(cosmic) in
equation~(\ref{eq:04}) is treacherous in view of possible systematic
errors. In principle it is not possible to give a reliable estimate of
systematic errors because they are due to the unknown. But one can
list points where one depends on assumptions and where random errors
of correction factors perpetuate as systematic errors.

     The value of $H_{0}$ and its systematic error depends on two
parameters, i.e.\ the intercept $C_{BVI}$ of the Hubble diagram of
SNe\,Ia and the absolute magnitude $M_{BVI}$ of the calibrating
SNe\,Ia. The systematic errors of these two parameters will be
discussed in turn.

     The intercept $C_{BVI}$ depends on the recession velocities $v$
and on the apparent maximum magnitudes $m_{BVI}$ of SNe\,Ia with
$3000<v<20\,000\kms$. Systematic errors of the observed velocities and
of their correction to the CMB frame are negligible. Equally the
observed apparent magnitudes do not introduce a systematic error. The
magnitudes are corrected for Galactic and internal absorption. The
internal absorption is based on the intrinsic colors $(B\!-\!V)$ and
$(V\!-\!I)$ of SNe\,Ia at $B$ maximum and 35 days thereafter and on
reddening-absorption ratios ${\cal R}_{BVI}$ specificly derived for
SNe\,Ia (\citeauthor{Reindl:etal:05}). These absorption corrections --
which remove any dependence of $C_{BVI}$ on galaxy type or on the size
of the absorption (\citeauthor{Reindl:etal:05}, Table~8, solutions
6-8) -- do not introduce a systematic error as long as they are
consistently applied to the Hubble diagram SNe\,Ia and to the
nearby calibrating SNe\,Ia. The same holds for the normalization of
the magnitudes to a decline rate of $\Delta m_{15}=1.1$ and a color of
$(B\!-\!V)^{\rm corr}=-0.024$ (\citeauthor{Reindl:etal:05},
equation~23). Therefore the value of the intercept 
$C_{BVI}$ as such does not introduce a systematic error.

     The absolute magnitudes $M_{BVI}$ of the calibrating SNe\,Ia may
carry systematic errors for two reasons: (1) errors of their apparent
magnitudes due to inconsistencies of their absorption correction and
of their normalization to $\Delta m_{15}=1.1$ and $(B\!-\!V)^{\rm
  corr}=-0.024$, and (2) metallicity-related distance errors and a
zero-point error of the adopted distance scale.

\noindent
1)a. While the apparent maximum magnitudes of the distant SNe\,Ia are
based on CCD photometry, four of the oldest calibrators have
photographic photometry, but the magnitudes were transformed into the
standard $B,V$ system for SN\,1937C  [\citealt{Schaefer:96a};
following this source we have discarded the fainter photometry of
\citet{Pierce:Jacoby:95}. If it were included here the mean luminosity
of the 10 calibrators would decrease by only $\la0.01\mag$],
1960F \citep{Tsvetkov:83,Schaefer:96b,Saha:etal:96b}, 
1974G \citep{Schaefer:98}, and 1981B \citep{Schaefer:95}. 
The photoelectric $U,B,V$ photometry \citep{Ardeberg:deGroot:73} of
the bright and far outlying SN\,1972 is not affected by background
light of the parent galaxy NGC\,5253 and is reliable. 

     The template-fitted lightcurve parameters of these five
SNe\,Ia have been compiled in \citeauthor{Reindl:etal:05}. They have
somewhat larger random errors than later SNe\,Ia and are given
correspondingly lower weights, but there is no reason why they should
introduce any {\em systematic\/} error. Figure~\ref{fig:02} shows that
their absolute magnitudes are randomly distributed in comparison to
later SNe\,Ia as to the year of discovery, distance, reddening etc. 
It would be arbitrary to leave out the older SNe\,Ia. In any case it
would make little difference (see \ref{sec:H0cosmic:previous:Riess}).

\noindent
b. The absorption corrections of the nearby and distant SNe\,Ia are
the product of the consistently derived color excesses $E(B\!-\!V)$
and the absorption factor ${\cal R}_{BVI}$. Since there is a mean
color excess difference between the two groups of SNe\,Ia of
$\Delta\langle E(B\!-\!V)\rangle =0.027$ (Table~\ref{tab:05}), an
error of the adopted ${\cal R}_{BVI}$ of as much as $\pm0.5$ will
cause a systematic error of the absorption $A_{BVI}$, and hence of the
absorption-corrected magnitudes of only $0.014\mag$. (Note that
variations of ${\cal R}_{BVI}$ from galaxy to galaxy are expected to
average out in first approximation for samples with $\sim\!10$ and
more elements). 

\noindent
c. In analogy to b. an error of the slope $a$ of the $M_{BVI}-\Delta
m_{15}$ relation will introduce a systematic error. The mean value of
$\Delta m_{15}$ of the calibrators and distant SNe\,Ia differs by
$\delta\Delta m_{15}=0.17$ (Table~\ref{tab:05}). If $a$ has a $10\%$
error of $\sim0.2$ (see \citeauthor{Reindl:etal:05}, Table~5) the
resulting systematic error will remain within $0.04\mag$.

\noindent
d. The normalization of the SNe\,Ia magnitudes to a common color
(\citeauthor{Reindl:etal:05}, equation~23) will not introduce an
additional systematic error because the calibrators and distant
SNe\,Ia have identical mean $(B\!-\!V)$ colors (Table~\ref{tab:05}).

\noindent
2)a. The present metallicity corrections $\Delta\mu_{Z}$, as defined
in equation~(10) of \citeauthor{Saha:etal:06}, were derived on purely
empirical grounds. They rest on two hypotheses: 
(1) The observed difference of the $P$-$L$
relations in the Galaxy and LMC is caused by the metal difference of
the two galaxies, and (2) the slopes of the $P$-$L$ relations transform
smoothly from LMC to the Galaxy as [O/H] (or $Z$) increases. Both
points are consistent with earlier models cited in \S~\ref{sec:intro},
but it remained the concern that the increase of $Y$ accompanying any
increase of $Z$ would counterbalance the shift. Only recent models by
\citet{Fiorentino:etal:02} and \citet{Marconi:etal:05} show that the
effect of changing $Y$ is small, and the Cepheids become progressively
cooler at $L=\mbox{const}$ as $Z$ increases over a wide range of
$\Delta Y/ \Delta Z$. Thus our basic assumption as to why the Galactic
$P$-$L$ relations differ from those of LMC is justified by theory.

     The observed $P$-$L$ relations in $V$ and $I$ of the Galaxy yield
{\em larger\/} distances for long-period Cepheids with $\log P\ga1.0$
than the observed LMC $P$-$L$ relations (\citeauthor{Saha:etal:06},
Fig.~8). Therefore the metallicity corrections $\Delta\mu_{Z}$ must be
positive as the metallicity increases.
The models of \citet{Marconi:etal:05} give positive values of
$\Delta\mu_{Z}$ only for very high metallicities of $Z\ga0.03$, but
quantitative agreement cannot be expected since $\Delta\mu_{Z}$ and
its sign depend on the subtle interplay of the slopes in $V$ and $I$
of the Galactic and LMC $P$-$L$ relations, which in turn depend on the
ridge lines of Cepheids in the $\log L - \log T_{\rm e}$ plane and
which do not necessarily coincide with the mid-line of the instability
strip because the evolutionary crossing times during the second
crossing of the strip are a strong function of temperature
\citep{Alibert:etal:99}. 

     The mean metallicity of the Cepheids in the ten calibrating
galaxies is smaller by only $\Delta[\mbox{O/H}]_{\rm Sakai}=0.05$ than
the adopted metallicity of the Galactic Cepheids. It follows from the
models of \citet{Marconi:etal:05} that in this case the Galactic
$P$-$L$ relations are a much better approximation than those of
LMC. In fact, the mean metallicity correction of $\mu^{0}$(Gal) of the
ten calibrating galaxies amounts to only $\Delta\mu_{Z}=0.022$. 
It is therefore likely that the metallicity corrections affect the
zero-point of the distance scale by $\le0.10\mag$. 

     That this estimate is realistic is further supported by the
comparison in \citeauthor{Saha:etal:06} of the metallicity-corrected
moduli $\mu^{0}_{Z}$ with independent TRGB and velocity distances,
by the (near) independence of the luminosity of the calibrating
SNe\,Ia on metallicity, and by the fact that the metal-rich and
metal-poor Cepheids in NGC\,5457 (M\,101), as published by
\citet{Kennicutt:etal:98}, yield the same distance with the present
metallicity corrections. 

    One proviso is added. In \citeauthor{Saha:etal:06} it was
explained that an average subsolar abundance was adopted for the
Galactic Cepheids ($\langle[\mbox{O/H}]\rangle=8.60$). A solar
abundance of $8.70$ is not completely excluded because the O lines are
weak \citep{Kovtyukh:etal:05}. In that case the slope of the
metallicity corrections would become shallower and the distances of 
the 10 parent galaxies of the calibrating SNe\,Ia would decrease by
$0.037\mag$ on average. 

\noindent
b. The adopted zero-point of the distance scale is a source of
systematic error. Actually the present distance scale depends on two
zero-points, one for the Galactic and one for the LMC $P$-$L$
relation. The Galactic P-L relation was calibrated in
\citeauthor{Sandage:etal:04} in equal parts by 33 Cepheids in open
clusters and by BBW distances of 36 Cepheids.
The clusters are fitted to the ZAMS of the Pleiades at
$\mu^{0}=5.61\mag$; this well determined value rests on several
determinations including the {\em trigonometric\/} HIPPARCOS parallax
\citep{Makarov:02}. It was discussed in \citeauthor{Saha:etal:06} that
the calibrating clusters have solar metallicity on average, thus the
ZAMS fitting is justified. The BBW luminosities taken from
\citet{Fouque:etal:03} and \citet{Barnes:etal:03} are fainter at $\log
P=1.5$ than those from clusters by $0.11$, $0.14$, and $0.21\mag$ in
$B$, $V$, and $I$. They may indeed be somewhat faint for arguments
discussed in \citeauthor{Sandage:etal:04}. They are also fainter by
$0.24\pm0.08\mag$ than the seven Cepheids with interferometric
diameter measurements \citep{Kervella:etal:04}. But averaging in these
additional determinations with their proper weights moves the adopted
Galactic zero-point by less than $0.10\mag$.

     The LMC zero-point rests on an adopted weighted modulus of
$(m-M)^{0}_{\rm LMC}=18.54$ as derived in
\citeauthor{Tammann:etal:03} from 13 determinations by various
authors, yet excluding distances based on the $P$-$L$ relation of
Cepheids. Eclipsing binaries give usually rather lower distances 
\citep[e.g.][]{Fitzpatrick:etal:03}, but the particularly well
determined distance of HV982 of $18.50\pm0.05$
\citep{Fitzpatrick:etal:02} is compatible with the adopted
modulus. 
It is also in good agreement with more recent determinations, i.e.\ 
$18.59\pm0.09$ from the TRGB \citep{Sakai:etal:04},
$18.52\pm0.03$ from $K$ magnitudes of RR\,Lyr stars based on Bono's
   et~al. (\citeyear{Bono:etal:03}) semi-theoretical zero-point
   \citep{Dall'Ora:etal:04}, 
$18.53\pm0.06$ from the BBW method \citep{Gieren:etal:05} and
$18.55\,(\pm0.10)$ from RR\,Lyr stars \citep{Sandage:Tammann:06a}.
It is therefore unlikely that the LMC zero-point is off by as much as
$0.10\mag$.

     The distances of the ten galaxies calibrating the SN\,Ia
luminosity are secured by the Galactic {\em and\/} LMC zero-points. 
The error of their combined weight must be smaller than $0.10\mag$. 

\noindent
c. In \citeauthor{Saha:etal:06} (\S~6 there) it was pointed out that
Cepheid distances are always dependent on period if the observed
slopes in $V$ and $I$ are not identical to the corresponding slopes of
the $P$-$L$ relations used for calibration. The dependence of the
adopted distances $\mu^{0}_{Z}$ on period was expressed by a
$\pi$-factor such that $\Delta \mu^{0}_{Z} =
\pi\cdot\Delta\log P$. The Cepheids of the ten calibrating galaxies
have an overall mean period of $\langle\log P\rangle=1.45$ (28 days)
and $\langle\pi\rangle = 0.40\pm0.32$ (Table~A1 in
\citeauthor{Saha:etal:06}). If future discoveries of fainter Cepheids
will decrease the mean period to say 20 days ($\langle\log
P\rangle=1.30$) the galaxy distances will decrease by $0.06\mag$ on
average. This estimate is an oversimplification because it is
foreseeable that additional Cepheids will change the observed slope of
the $P$-$L$ relations and hence also the value of $\pi$, but it serves
to illustrate the uncertainty inherent in Cepheid distances.
 
     Adding the various error sources in quadrature gives a total
systematic error of $0.17\mag$. The corresponding error on $H_{0}$ is
$\Delta H_{0}=5.0$.

\subsection{Previous Determinations of \boldmath{$H_{0}$} 
            with SNe\,Ia as Standard Candles}
\label{sec:H0cosmic:previous}
Over the past 24 years a number of attempts have been made to use
SNe\,Ia as standard candles and to derive -- after their luminosity
has been calibrated in a few nearby cases -- the large-scale (or not
so large-scale) value of $H_{0}$. Table~\ref{tab:06} lists 24 original
papers which are devoted to this aim. 

     It is amusing to note that the overall mean of the 24
determinations of $H_{0}$ in Table~\ref{tab:06} is $63.5\pm1.5$, i.e.\
very close to the present result. However, the seeming agreement is
fortuitous, because the individual determinations are based on
different $P$-$L$ relations, different calibrators, different absorption
corrections, and on different metallicity and decline rate
corrections, if any such corrections are applied at all. The wide
variation of $H_{0}$ in Table~\ref{tab:06} is therefore mainly due to
systematic effects. It is, however, noteworthy that over the last 20
years all values of $H_{0}$ agree within their quoted errors with only
three or four exceptions.

     The values of our team (Saha et~al.\ 1994-2001;
\citealt{Parodi:etal:00}; present paper) have increased over the
years from 52 to 62. The lowest value had still to rely on a uncertain
calibration through the brightest stars in only two parent galaxies,
and one of the calibrators, SN\,1954a, turned later out to belong to
the overluminous class of which SN\,1991T is the prototype. Also
SN\,1895B, which was used in the first papers, may belong to this
class because it is $0.25\mag$ brighter than SN\,1972E in the same
galaxy (NGC\,5253). Two other eventually discovered effects went also
into the direction of increasing $H_{0}$. The decline rate correction,
first quantified by \citet{Phillips:93}, increases (in its present
form) $H_{0}$ by 3 units, and the passage from a $\Omega_{\rm M}=1$
universe to $\Omega_{\rm M}=0.3$, $\Omega_{\Lambda}=0.7$ brought an
additional increase of 0.8 units. 

     The latest increase of $H_{0}$ by 6.5\% over our value in
\citet{Saha:etal:01} is due to an accumulation of small effects. It
should first be noted that the values $M_{BVI}$ and $C_{BVI}$ here are
reduced to $\Delta m_{15}=1.1$ and $(B\!-\!V)^{\rm corr}=-0.024$,
while the reference values were $\Delta m_{15}=1.2$ and
$(B\!-\!V)^{\rm corr}=-0.01$ in 2001. 
This affects {\em equally\/} the apparent magnitudes $m_{BVI}^{\rm
  corr}$  of the calibrators and the intercepts $C_{BVI}$ of the
distant SNe\,Ia, and has no effect on $H_{0}$. However, the reduction
of the absorption factor ${\cal R}_{BVI}$ for the dust absorption of
SNe\,Ia in the host galaxy of originally $4.3$, $3.3$, and $2.0$ to
the present values $3.65$, $2.65$, and $1.35$ makes the more highly
reddened calibrators (see Table~\ref{tab:05}) fainter by $0.03\mag$
relative to the distant SNe\,Ia, which now increases $H_{0}$ by 1.5\%. 
Also the coefficients $a_{BVI}$ of the $\Delta m_{15}$ correction had
to be increased for reasons explained in \citeauthor{Reindl:etal:05};
this dims the calibrators with their higher $\langle\Delta
m_{15}\rangle$ (see Table~\ref{tab:05}) more than the distant SNe\,Ia
by $0.02\mag$ on average (1\%).
A further increase comes from the new photometry of SN\,1998aq (1\%)
and the arrival of the slightly faint tenth calibrator, SN\,1998ae,
(0.6\%). Finally the adopted Cepheid distances $\mu_{\rm Z}^{0}$ of
the calibrating host galaxies, as derived in \citeauthor{Saha:etal:06}
from new $P$-$L$ relations and new metallicity corrections and used in
Table~\ref{tab:01}, are 1\% smaller than those in
\citet{Saha:etal:01}, the latter containing a bulk metallicity
correction of $0.06\mag$ and resting on a slightly brighter zero-point
of $(m-M)_{\rm LMC}^{0}=18.56$. These effects together increase our
2001-value of $58.7\pm2.0$ to $61.7\pm2.1$, which is statistically the
same as the present value of $62.3\pm1.3$.

\subsubsection{Comparison of $H_{0}$ with \citet{Freedman:etal:01}} 
\label{sec:H0cosmic:previous:NASA}
One of the strongly deviating $H_{0}$ values in
Table~\ref{tab:06} is from \citet{Freedman:etal:01}, whose value of
$H_{0}=72$ has been widely adopted. These authors
have derived the Cepheid distances of the host galaxies of their six
calibrating SNe\,Ia from the single-fit LMC $P$-$L$ relations of
\citet{Udalski:etal:99}, a zero-point of $(m-M)^{0}_{\rm LMC}=18.50$,
and including a metallicity correction close to
\citet{Kennicutt:etal:98}. These $P$-$L$ relations are now untenable
(\citeauthor{Sandage:etal:04}). If one applies the $P$-$L$ relations
of LMC of \citeauthor{Sandage:etal:04}, with their break at
$P=10\;$days, to the same six SN\,Ia-calibrating galaxies their moduli
increase by $0.17\mag$, or -- after application of the same
metallicity correction -- by even $0.23\mag$. Further, with the adopted
moduli $\mu^{0}_{\rm Z}$ from Table~\ref{tab:01}, column~(12), which
contain the period-dependent metallicity correction of
\citeauthor{Saha:etal:06}, the discrepancy increases to
$0.35\mag$. With our moduli \citet{Freedman:etal:01} would have
obtained $H_{0}=60.4$. Thus their very high value of $H_{0}$ is solely
caused by their (too) small distances compared with ours. 
In particular it is {\em not\/} due to differences of the HST-based
Cepheid magnitudes, because the Freedman team have reproduced them in
seven of our program galaxies to within a few $0.01\mag$
\citep[][Table~3]{Gibson:etal:00}. --
The almost equally high value of $H_{0}$ by \citet{Altavilla:etal:04}
is based on eight calibrators of \citet{Saha:etal:01} plus the
overluminous SN\,1991T, but their result is not independent of
\citet{Freedman:etal:01} because they have adopted the Cepheid
distances of the latter source. In addition they have applied two
versions of a theoretical metallicity correction which changes sign
about at the metallicity of Galactic Cepheids, independent of
period. Their final moduli are smaller than ours on average by $0.33$
to $0.43\mag$. Using our moduli, they would have obtained
$H_{0}=58\!-\!61$.  
It is no surprise that \citet{Wang:etal:06} found for $H_{0}$ the same
value as \citet{Freedman:etal:01} because they used for their
calibrators the Cepheid distances of the latter source.
 
\subsubsection{Comparison of $H_{0}$ with \citet{Riess:etal:05}} 
\label{sec:H0cosmic:previous:Riess}
The most deviating value of $H_{0}$ in Table~\ref{tab:06} comes
from \citet{Riess:etal:05}. It is based on only four calibrators whose
Cepheid distances were derived from the LMC $P$-$L$ relations of
\citet{Thim:etal:03}, which are a slightly earlier version of those in
\citeauthor{Sandage:etal:04}, but with $(m-M)^{0}=18.50$ and including
a metallicity correction from \citet{Sakai:etal:04}. The mean
metallicity of the four calibrators is very close to the Galactic
value. Hence it would have been a more obvious choice to use the {\em
  Galactic\/} $P$-$L$ relation of \citeauthor{Sandage:etal:04}. 
In that case the authors would have found $H_{0}=63.1$ instead of
$H_{0}=73$. This is close to the value $63.3\pm1.9$ from the present
metallicity corrected moduli $\mu^{0}_{\rm Z}$ of the four
calibrators. \citet{Riess:etal:05} have excluded six calibrators as
being too old or too heavily absorbed, yet we find for these six
SNe\,Ia $H_{0}=61.0\pm2.0$, which is the same within statistics as for
their four calibrators. Clearly, the authors' result is not due to the
choice or treatment of their calibrating SNe\,Ia, {\em but only to the
  fact that they apply a $P$-$L$ relation for metal-poor Cepheids to a
  sample of metal-rich Cepheids}. Their correction for metallicity is
insufficient, 
particularly because their calibrating Cepheids have exceptionally
long periods ($\langle P\rangle=35\;$days) and require large
corrections.

\section{THE LOCAL VALUE OF \boldmath{$H_{0}$}}
\label{sec:H0local}
The global value of $H_{0}$ in \S~\ref{sec:H0cosmic} considered
only SNe\,Ia in the velocity range $3000<v<20\,000$ (or
$30\,000$)$\;\kms$. It remains the interesting question as to the mean
``local'' value of $H_{0}$, say within $2000\kms$. This question is
persued in the following sections using Cepheid distances, local
SNe\,Ia, the mean cluster distances of Virgo and Fornax, as well as
21cm-line width and TRGB distances.

     Since small distances, down to $\sim\!2\;$Mpc, and small
velocities are considered here, care is taken to correct them
appropriately. All distances in this paragraph refer to the barycenter
of the Local Group, assumed at 2/3 (0.54$\;$Mpc) of the distance
towards M\,31 \citep{Sandage:86}; these distances are denoted with
$r_{0}$ or $\mu^{0}_{0}$.

     The heliocentric velocities $v_{\odot}$ of the galaxies are
corrected for the solar motion with respect to the barycenter of the
Local Group following \citet{Yahil:etal:77}. Similar solar-motion
solutions by \citet{Sandage:86} and \citet{Richter:etal:87} lead to
slightly larger scatter of the 
Hubble diagram of nearby galaxies. The more deviating solution by
\citet{Karachentsev:Makarov:01} causes still larger Hubble scatter;
the solution may be influenced by orbital velocities of companion
galaxies moving about a larger galaxy. The solar-motion-corrected
velocities $v_{0}$ are then reduced to $v_{00}$, which are the
velocities as seen from the barycenter of the Local Group. This
(small) correction is only viable on the assumption that the observed
recession velocities are strictly radially away from the
barycenter. Finally the velocities $v_{00}$ are corrected for a
self-consistent Virgocentric infall model with a local infall vector
of ${\vec v} =220\kms$ and an adopted $R^{-2}$ density profile of
the Virgo complex
\citep{Yahil:etal:80,Tammann:Sandage:85,Kraan-Korteweg:86}, where
$v_{220}=v_{00}+\Delta v_{220}$.
These authors have calculated $\Delta v_{220}$ without knowledge of
the galaxy distances from $v_{0}$, ${\vec v}$, and the mean velocity
of the Virgo cluster $\langle v_{0}\rangle$. In the present case,
where all galaxy distances are known, it is easier to use 
\begin{equation}
 \Delta v_{220} = 220 (\cos \alpha +
 \frac{r_{0}(\mbox{Virgo})}{R(\mbox{galaxy})} \cdot \cos \beta) \kms,
\label{eq:05}
\end{equation}
where $r_{0}(\mbox{Virgo})$ is the distance of the Virgo cluster,
$R(\mbox{galaxy})$ is the distance of the galaxy from M\,87, 
$\alpha$ is the angle of the galaxy away from M\,87 as
seen from the barycenter, and $\beta$ is the angle of the barycenter
away from M\,87 as seen from the galaxy. The route through
equation~(\ref{eq:05}) is here actually preferable because the errors
of the small distances $r_{0}$ are small compared to the errors of
$v_{0}$, which may be contaminated by important peculiar velocities.

\subsection{\boldmath{$H_{0}$}(local) from Cepheids}
\label{sec:H0local:cepheid}
\begin{figure*}[p] 
   \epsscale{1.0}
   \plotone{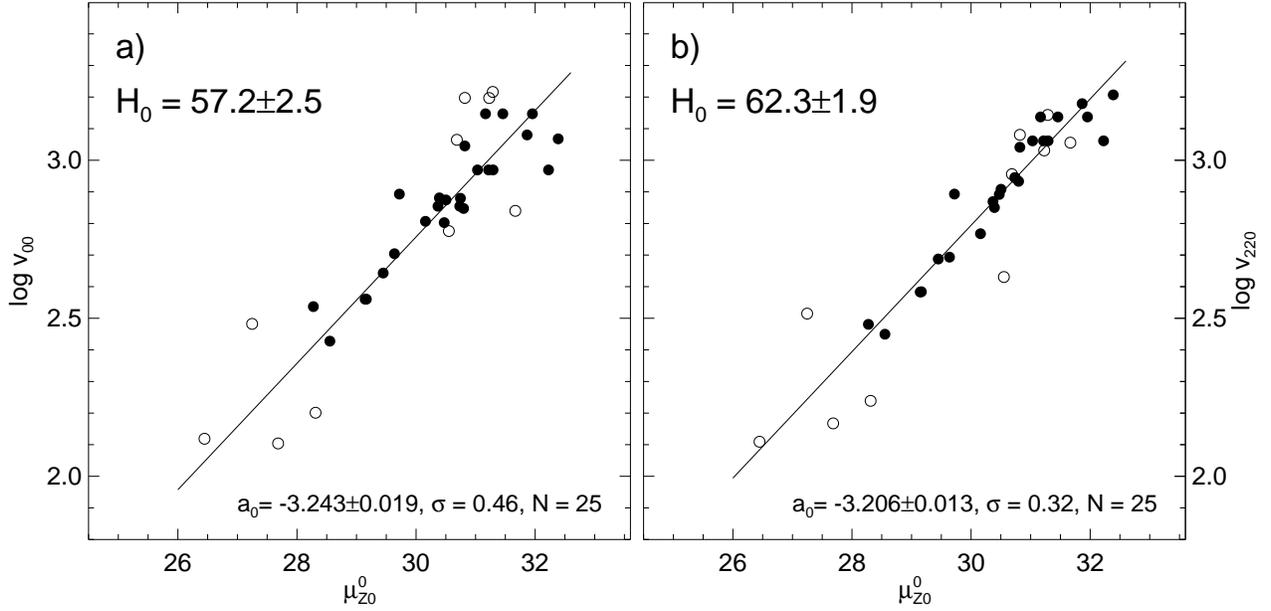}
   \caption{The distance-calibrated Hubble diagram using 18 galaxies
   and 7 Virgo or Fornax cluster members with Cepheid distances
   (filled symbols). Galaxies within  $25^{\circ}$ from the Virgo
   cluster (M\,87), but outside the cluster, and with  $\mu_{{\rm
   Z}0}^{0}<28.2$ are shown as open symbols. a) using velocities
   $v_{00}$ reduced to the gravicenter of the Local Group. 
   b) using velocities $v_{220}$ corrected in addition for
   Virgocentric infall. The fitted Hubble lines $\log v=0.2\mu + 
   a_{0}$ are only through the filled symbols.}  
\label{fig:04}
\end{figure*}

The velocity field within $2000\kms$ is mapped in Figure~\ref{fig:04}
by 25 galaxies, including four members of the Virgo cluster and three
members of the Fornax cluster, whose Cepheid distances are taken from
\citeauthor{Saha:etal:06} (Table~A1). The cluster members are plotted
with the mean cluster velocity. Also shown as open symbols are four
galaxies with $\mu^{0}_{{\rm Z}0}<28.2$ and six galaxies, which are not
members of the Virgo cluster, but with angular distances from the
cluster smaller than $\alpha_{{\rm M}87}<25^{\circ}$. The four nearby
galaxies are not used for the solution because the contribution of
their peculiar velocities may be important. The six field galaxies
with $\alpha_{{\rm M}87}<25^{\circ}$ will be discussed below.

     The distances $\mu^{0}_{{\rm Z}0}$ and velocities $v_{00}$, all
reduced to the Local Group barycenter, of the 25 galaxies give a
small Hubble constant ($57.2\pm2.5$) and large scatter
($\sigma_{m}=0.46\mag$), (Fig.~\ref{fig:04}a). If, however, the
velocities are corrected for Virgocentric infall the scatter is
significantly reduced to $\sigma_{m}=0.32\mag$ and $H_{0}$(local)
becomes $62.3\pm1.9$ (Fig.~\ref{fig:04}b).
\clearpage
\clearpage

\subsection{\boldmath{$H_{0}$}(local) from nearby SNe\,Ia}
\label{sec:H0local:sneia}
\begin{figure*}[p] 
   \epsscale{1.0}
   \plotone{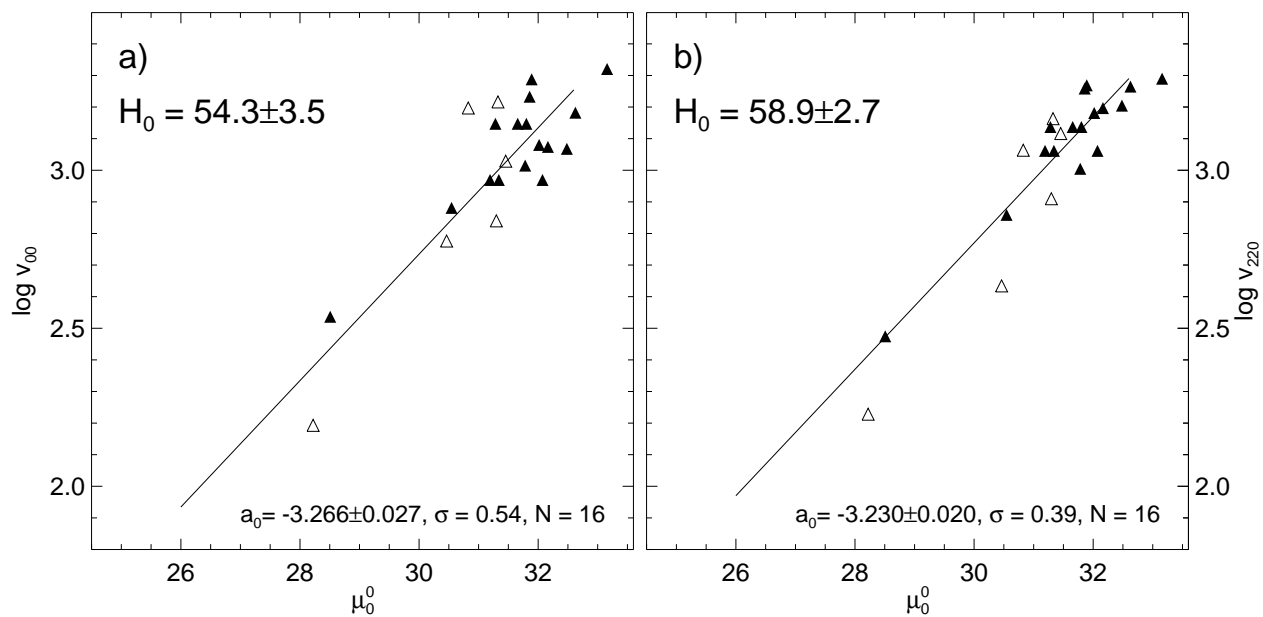}
   \caption{The distance-calibrated Hubble diagram of SNe\,Ia with
   $v_{220}<2000\kms$. Those outside the Virgo cluster but within
   $25^{\circ}$ from the cluster center (M\,87) and those with
   $\mu_{0}^{0}<28.2$ are shown as open symbols. a) using velocities
   $v_{00}$ reduced to the gravicenter of the Local Group. b) using
   velocities $v_{220}$ corrected for Virgocentric infall. The fitted
   Hubble lines are only through the filled symbols.} 
\label{fig:05}
\end{figure*}

Fully corrected apparent $V$ magnitudes at maximum are given in
\citeauthor{Reindl:etal:05} for 16 SNe\,Ia with $v_{0}<2000\kms$. Their
magnitudes are combined with the weighted absolute magnitude of
$M_{V}=-19.46$ from Table~\ref{tab:03} to obtain distance moduli
$\mu^{0}$, which are transformed, as before, to distances
$\mu^{0}_{0}$ from the Local Group barycenter. The latter are plotted
in a Hubble diagram in Figure~\ref{fig:05} (filled symbols). The
velocities $v_{00}$, again corrected to the barycenter, give large
scatter ($\sigma_{m}=0.54\mag$) and a small value of
$H_{0}=54.3\pm3.5$ (Fig.~\ref{fig:05}a), but after a correction for
Virgocentric infall these numbers become $\sigma_{m}=0.39\mag$ and
$H_{0}=58.9\pm2.7$ (Fig.~\ref{fig:05}b), i.e.\ the value of $H_{0}$ is
statistically the same as on large scales (equation~\ref{eq:04}).

     One nearby SN\,Ia with $\mu^{0}_{0}<28.2$ and six SNe\,Ia within
the $25^{\circ}$ ring about the Virgo cluster have not been used for
the solution for the same reason given in
\S~\ref{sec:H0local:cepheid}. They are shown as open symbols in
Figure~\ref{fig:05}. 

\begin{figure*}[p] 
   \epsscale{1.0}
   \plotone{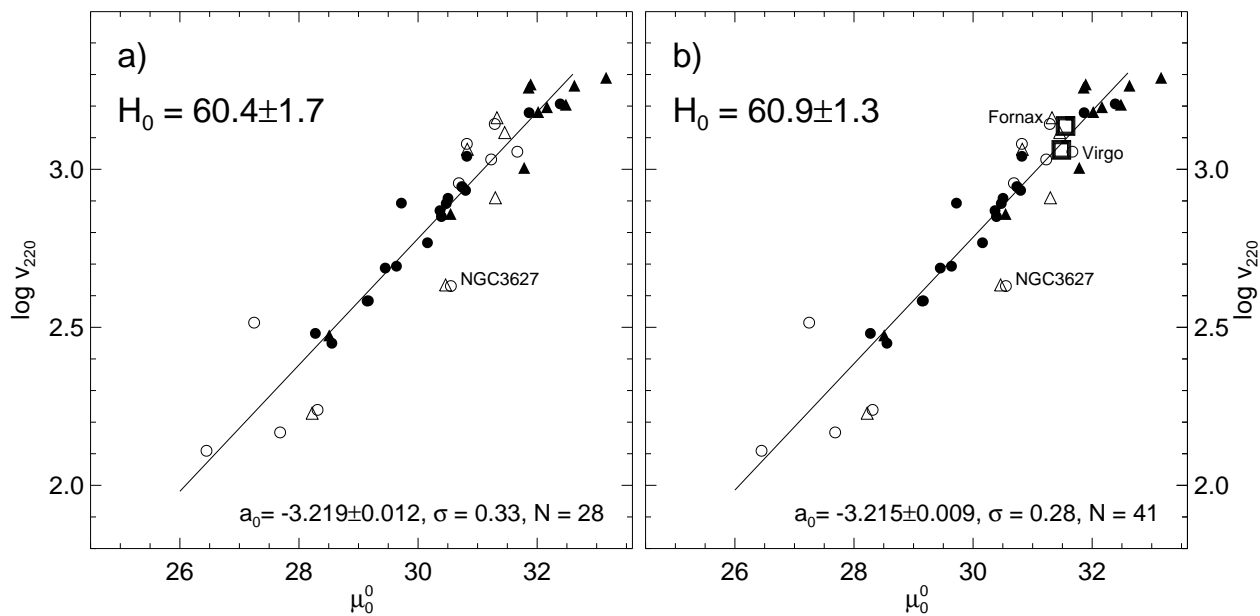} 
   \caption{The combined distance-calibrated Hubble diagram of
   field galaxies with Cepheid distances (circles) and of SNe\,Ia
   (triangles). Objects inside $25^{\circ}$ from the Virgo cluster
   (M\,87) and those with $(m-M)^{0}<28.2$ are shown as open
   symbols. The Hubble line is fitted to only the filled symbols. b)
   the Virgo and Fornax clusters are added at their mean distance
   and mean velocity (from Table~\ref{tab:07}).}
\label{fig:06}
\end{figure*}

     The data of the Cepheids in Figure~\ref{fig:04} and of the local
SNe\,Ia in Figure~\ref{fig:05} are combined in the Hubble diagram of
Figure~\ref{fig:06}, where only $v_{220}$ velocities are shown. The 28
high-weight distances of field galaxies are plotted, as before, as
filled symbols. They define a Hubble line corresponding to
$H_{0}=60.4\pm1.7$ and $\sigma_{m}=0.33\mag$)
(Fig.~\ref{fig:06}a). This result is only slightly changed by adding
the 13 Virgo and Fornax cluster members from
\S~\ref{sec:H0local:distance} to become 
\begin{equation}
   H_{0}=60.9\pm1.3, \quad\ \sigma_{m}=0.28\mag,
\label{eq:06}
\end{equation}
which we adopt for the local value. We emphasize again the agreement
with the global value. 
\clearpage
\clearpage
\clearpage
\clearpage

     Objects with $\mu^{0}_{0}<28.2$ are shown as open symbols in
Figure~\ref{fig:06}. They are not used for equation~(\ref{eq:06}),
although their inclusion would not change $H_{0}$, but it would
increase the scatter (due to the {\em relatively\/} large peculiar
motions). 

     The 15 galaxies with distances from Cepheids or SNe\,Ia, which
are shown as open symbols in the upper part of Figure~\ref{fig:06},
lie in a ring of radius $25^{\circ}$ about the Virgo cluster. 
Their separate treatment here is a precaution because
it may be suspected that the surroundings of a cluster have
particularly large peculiar velocities. This seems supported by a
Hubble diagram plotting $\log v_{00}$ versus $\mu^{0}_{0}$, where the
scatter is as high as $\sigma_{m}=0.72\mag$. Even if the deviating
galaxy NGC\,3627 is excluded, the scatter remains high at
$\sigma_{m}=0.45\mag$. However, if the velocities are corrected for
Virgocentric infall the scatter reduces to $\sigma_{m}=0.31$ which is
even less than $\sigma_{m}=0.33$ from the field galaxies outside the
$25^{\circ}$ ring. Two conclusions follow:  
(1) the region about the Virgo cluster is hardly more turbulent than
the general field, and (2) the Virgocentric infall model, which
already efficiently reduced the velocity scatter in
Figure~\ref{fig:04} and \ref{fig:05}, is surprisingly successful even
in regions close to the cluster.

     The agreement of $H_{0}(\mbox{local})=60.9\pm1.3$ and
$H_{0}(\mbox{global})=62.3\pm1.3$ does not mean that the Hubble flow
is unaltered by gravity, because it must be stressed again that a
selfconsistent Virgocentric infall model with a local infall vector of
$220\kms$ has been subtracted from the observed velocities to
compensate for the excess gravity of the cluster. 

   An interesting by-product of Figure~\ref{fig:06}a is its small
scatter of $\sigma_{\mu}=0.33\mag$, which is reduced to $0.28\mag$ if
an average error of $0.15\mag$ of the distance determinations is
allowed for. This implies that peculiar motions contribute to the
scatter with only $\partial v /v =0.14$, and hence that in the
distance interval of $500\la v\le 1500\kms$ the peculiar velocities of
field galaxies are restricted to $70-210\kms$.

\subsection{\boldmath{$H_{0}$}(local) from the Distance of the Virgo
  and Fornax Clusters} 
\label{sec:H0local:distance}
Cepheid and SNe\,Ia distances of Virgo and Fornax cluster members
render themselves for a distance determination of the two
clusters. The relevant data are shown in Table~\ref{tab:07}, where the
Cepheid distance moduli are taken from \citeauthor{Saha:etal:06}
(Table~A1, col.~[9]). The SN\,Ia moduli are the difference between the
fully corrected apparent magnitudes $m_{V}$ from
\citeauthor{Reindl:etal:05} (Table~2, col.~[9]) and the absolute
magnitude $M_{V}=-19.46$ (from Table~\ref{tab:03}, col.~[12]). The
21cm-line width distance of the Virgo cluster is from below
(\S~\ref{sec:H0local:21cm}). 

   In the case of the Virgo cluster the two SNe\,Ia 1960F (NGC\,4496A)
and 1981B (NGC\,4536) are omitted, because they do not lie in the cluster
proper, but in the complex W-cloud \citep{Binggeli:etal:93}. 
Also SN\,1999cl (NGC\,4501) is omitted because of its large absorption
(\citeauthor{Reindl:etal:05}). 
However, SN\,1990N (NGC\,4639) is included in Table~\ref{tab:07} as a
bona fide member of the Virgo cluster in spite of its large distance.
Its recession velocity agrees almost exactly with the cluster
mean. The galaxy has gravitationally interacted with NGC\,4654, which
shows additional effects of ram pressure from the cluster X-ray gas
\citep{Vollmer:03}. 
Moreover, if NGC\,4639 was a field galaxy in the cluster background its
distance and $H_{0}=62$ would require a recession velocity 
higher than observed by $\sim\!500\kms$. 
A peculiar velocity of this size of NGC\,4639 {\em and\/} NGC\,4654
would be very unusual for field galaxies. 

     While NGC\,4639 clearly lies on the far side of the Virgo
cluster, the three remaining Virgo galaxies with Cepheid distances
(NGC\,4321, 4535, and 4548) are on the near side,
because they have been selected for the HST observations on the basis
of their above average resolution \citep{Sandage:Bedke:88}. The effect
of this bias has been neglected by \citet{Freedman:etal:01} and
others. High resolution galaxies do indeed favor an incorrect small
mean distance of the Virgo cluster if used alone
\citep[][Fig.~7]{Tammann:etal:02}. The difficulty to determine the
exact Virgo cluster distance lies in the fact that its extent in depth
($\sim\!10\;$Mpc) is significantly larger than its projected radius.

     Table~\ref{tab:07} lists also a Virgo cluster distance from the
21cm-line width method (see \S~\ref{sec:H0local:21cm} below). The
cluster distance modulus of $31.65\pm0.08$ $(\sigma_{\mu}=0.59)$ is
derived from a {\em complete\/} sample of 49 inclined Virgo cluster
spirals, as compiled by \citet{Federspiel:etal:98}, and from the
Cepheid-based calibration of the method in equation~(\ref{eq:07})
below. The distance, however, has been reduced by $0.07\mag$ for the
fact that the cluster members at a given line width are redder in
$(B\!-\!I)$ on average and also HI-deficient if compared with the
calibrating field galaxies
\citep[see][\S~8]{Federspiel:etal:98}. As a consequence our
adopted TF distance of the Virgo cluster becomes
$\mu_{0}=31.58\pm0.08$. 
[See also \citet{Sandage:Tammann:06b} for a less restricted sample of
Virgo cluster galaxies giving $\mu_{0}=31.60\pm0.09$].

     The unweighted distances of the Virgo and Fornax cluster are
shown in Table~\ref{tab:07} as well as their velocities. The adopted
velocity of the Virgo cluster of $1165\kms$ is the mean of $v_{220}$
and the {\em independent\/} value of $v_{\rm cosmic}$ from 
\citet{Jerjen:Tammann:93}. The ensuing values of $H_{0}$ are
$58.1\pm4.6$ from Virgo and $66.8\pm4.0$ from Fornax.
To emphasize, our distances to the Virgo
and Fornax cluster are $\sim\!0.7\mag$ more remote than derived by
\citet{Freedman:etal:01}, signalling the $\sim\!14\%$ difference in
our respective values of $H_{0}$.

\subsection{\boldmath{$H_{0}$}(local) from 21cm-Line Widths}
\label{sec:H0local:21cm}
Several global properties of galaxies correlate with the galaxian
luminosity or diameter, e.g.\ the morphological luminosity classes of
van den Bergh \citep[e.g.][]{vandenBergh:60a,vandenBergh:60b,Sandage:99}, 
21cm-line widths \citep[TF; e.g.][]{Tully:Fisher:77,Sakai:etal:00} 
of spirals or the surface brightness fluctuations 
\citep[SBF; e.g.][]{Tonry:Schneider:88,Tonry:etal:00} 
and velocity dispersion-diameter relation 
\citep[D$_{n}-\sigma$ or ``fundamental plane''; 
e.g.][]{Faber:Jackson:76,Dressler:etal:87,Djorgovski:Davis:87,Kelson:etal:00} 
of early type galaxies. Their
zero-point calibration depends directly or indirectly on Cepheids, and
they are therefore sensitive to any change of the distance scale of
Cepheids. The disadvantage of these methods is their large intrinsic
scatter ($\sigma_{M}>0.3\mag$), which makes them, if applied to {\em
  apparent-magnitude-limited\/} (and often even to incomplete
distance-limited) galaxy samples, susceptible to observational
selection bias (of which Malmquist is an example), {\em leading always
  to too high values of $H_{0}$}. For large samples, methods have been
devised to compensate for such biases in first approximation
\citep[][]{Bottinelli:etal:88,Federspiel:etal:94,Sandage:94,Sandage:96,Teerikorpi:87,Teerikorpi:90,Teerikorpi:97}.

   One of the few examples of an (almost) {\em complete
distance-limited\/} sample, which is immune to observational selection
bias, has been compiled by \citet{Federspiel:99} for spirals with
inclination $i>45^{\circ}$ and $v_{0}\le1000\kms$. 
The apparent magnitudes $B_{T}$ of the 114 sample galaxies are taken
from the RC3 \citep{deVaucouleurs:etal:91} or, if not available, the
apparent magnitudes $m_{B}$ are used as listed in the NASA/IPAC
Extragalactic Database (NED, http://nedwww.ipac.caltech.edu). Most of the  
43 $m_{B}$'s come originally also from the RC3 and are in the same
system as the $B_{T}$'s. The magnitudes are corrected for Galactic
absorption following \citet{Schlegel:etal:98} and for the inclination
dependent total internal absorption, which are determined as described
in the Introduction to the RSA. The necessary galaxian axis ratios
$a/b$ are taken from the RC3. The line widths $w_{20}$ are taken from
the same source where available. In 15 cases they are the mean of all
$\log w_{20}$ values given in the Lyon Database for physics of
galaxies (HyperLeda, http://leda.univ-lyon1.fr). 
Many galaxies with line widths in both catalogs reveal a systematic
inclination-dependant difference of 
\begin{equation}
  \Delta\log w_{20} = -0.003i + 0.203
\label{eq:10}
\end{equation}
in the sense of RC3 $-$ HyperLeda. The additional line widths have
been reduced to the system of the RC3 by means of
equation~(\ref{eq:10}). 

\begin{figure}[p] 
   \epsscale{0.55}
   \plotone{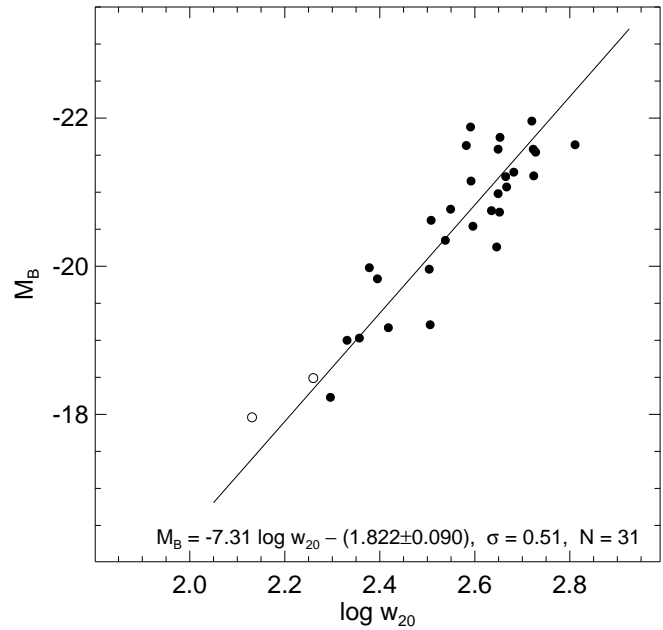}
   \caption{The calibration of the 21cm-line width $-$ absolute
   magnitude relation by means of 31 inclined galaxies with known
   Cepheid distances. The companion galaxies NGC\,5204 and 5585,
   shown as open symbols, are assumed at the same distance as
   M\,101. The scatter is much larger than can be accounted for by
   the error of the Cepheid distances.}
\label{fig:07}
\end{figure}

\begin{figure}[p] 
   \epsscale{0.55}
   \plotone{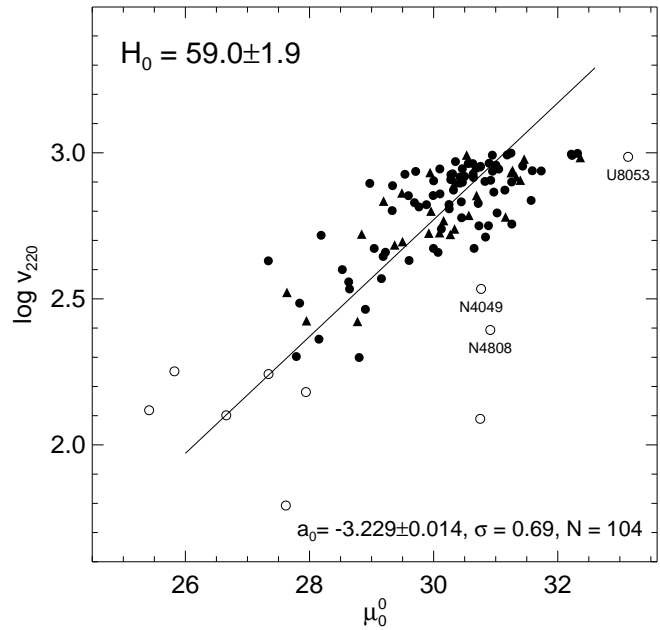}
   \caption{The distance-calibrated Hubble diagram of a complete
   sample of 114 inclined spirals with Tully-Fisher distances
   and with $v_{220}<1000\kms$. Galaxies with $\log w_{20}\ge2.4$ are
   shown as triangles, galaxies with $\log w_{20}<2.4$ as dots.    
   Galaxies with $v_{220}<200\kms$ and three deviating galaxies are
   shown as open symbols; they are not used for the solution. Note the
   large scatter. Member galaxies of the Virgo cluster are not shown.}
\label{fig:08}
\end{figure}

   There are 31 galaxies with known apparent magnitudes $B_{T}$ and
line widths $w_{20}$ from the RC3 for which Cepheid distances are available
from \citeauthor{Saha:etal:06} (Table~A1). After correction for
Galactic and internal absorption and in the case of $w_{20}$ for
inclination they define the calibration of the TF relation 
(Fig.~\ref{fig:07}):
\begin{equation}
   M_{B}=-7.31 \log w_{20} - (1.822\pm0.090), \sigma =0.51.
\label{eq:07}
\end{equation}
The slope of $-7.31$ was determined from Virgo cluster members by
\citet{Federspiel:etal:98} making allowance for errors of the apparent
magnitudes and of $\log w_{20}$ and for the depth effect of the
cluster. However, the slope is not well determined. Changing the
assumptions on the errors, or determinating the slope from the
Cepheid-calibrated galaxies (in spite of their restricted range in
$\log w_{20}$) can change the slope by a full unit in either
direction. Fortunately this affects the {\em mean\/} cluster distance
by only a few $0.01\mag$. The 31 calibrators give a scatter of the TF
relation of $\sigma_{MB}=0.51\, (\pm0.09)\mag$, which is
(insignificantly) larger than $\sigma_{MB}=0.43\, (\pm0.10)\mag$ found
by \citet{Sakai:etal:00} from 21 calibrators.
If equation~(\ref{eq:07}) is applied to the 111 galaxies in the
$1000\kms$ sample one obtains their distances, which are plotted with
their corresponding recession velocities ($\log v_{220}$) in
Figure~\ref{fig:08}. 
Excluding galaxies with $v_{220}<200\kms$ and three
strongly deviating galaxies, one obtains an intercept of the Hubble
line of $a_{0}=-3.229\pm0.014$, which corresponds to
$H_{0}=59.0\pm1.9$. The solution is remarkably robust. The galaxies
with $\log w_{20}\lessgtr2.4$ yield $H_0=59.3$ and 58.2,
respectively. The 16 field galaxies inside the $25^{\circ}$ about the
Virgo cluster yield, again in statistical agreement, $H_{0}=60.6$
(cf. \S~\ref{sec:H0local:sneia}). The scatter of $\sigma_{\mu}=0.69\mag$ in
Figure~\ref{fig:08} (or $\sigma_{\mu}=0.63\mag$ for $\log w_{20}>2.4$)
is surprisingly large, i.e.\ even larger than for the
spirals in the Virgo cluster ($\sigma_{m}=0.59\mag$) in spite of its
depth effect. Yet, if allowance is made for a scatter of $\sim\!0.3$
due to peculiar motions, the TF scatter in $B$ is reduced to
$\sigma_{\mu}=0.62\mag$ (or $0.55\mag$ for fast rotators).  
The still rather large scatter is not due to the mixture of $B_{T}$
and $m_{B}$ magnitudes because the scatter of the latter is only
insignificantly larger. Errors of other observational parameters
(inclination, internal absorption, line width) particularly of the
fainter sample members may blow up the scatter, but just the faintest
galaxies are essential for a complete distance-limited
sample. This illustrates the high price paid in using samples with
large $\sigma_{M}$ due to the always present observational selection
biases. 
In principle it would advantageous to use $I$ magnitudes for the TF
method, because the corrections for internal absorption are smaller
here but the paucity of available standard  $I$ magnitudes prevents
the definition of complete, {\em distance-limited\/} samples.

     Notwithstanding these difficulties, the mean value within
$1000\kms$ of $H_{0}=59.0\pm1.9$ from the TF relation for 104 field
galaxies agrees with the evidence from the smaller
Cepheid and SN\,Ia samples. Of course, this statement must be
relativated because the results are not independent since the TF
relation has been calibrated using the Cepheid distances. 

\clearpage
\clearpage
\clearpage
\clearpage

\subsection{\boldmath{$H_{0}$}(local) from TRGB Distances}
\label{sec:H0local:TRGB}
TRGB distances are of fundamental importance because they rest
entirely on Pop.~II stars and are hence {\em independent of any
Cepheid distances}. The brightness of the TRGB is based on globular
clusters \citep{DaCosta:Armandroff:90,Lee:etal:93,Bellazzini:etal:01};
its $I$-magnitude is, according to theoretical models
\citep{Cassisi:Salaris:97,Sakai:etal:04}, only moderately dependent on
metallicity. The disadvantage of the method is that its range is
restricted, even with HST, to $\sim\!10\;$Mpc.  

     In \citeauthor{Saha:etal:06} the TRGB distances of nine galaxies
by \citet{Sakai:etal:04}, for which also Cepheid distances are
available, were already used to confirm the zero-point of the Cepheid
distance scale to within $\sim\!0.1\mag$ and to demonstrate that the
adopted Cepheid distances carry no noticeable metallicity effect.

     Here many additional TRGB distances are used to determine a
very local value of $H_{0}$ and to compare it with the {\em
  independent\/} evidence from 
\S~\ref{sec:H0local:cepheid}$\;-\;$\ref{sec:H0local:21cm}.

     \citet{Karachentsev:etal:04,Karachentsev:etal:05} have determined
TRGB distances with HST for well over 100 galaxies of all types,
including many dwarf galaxies. 
Excluding as before galaxies with $\mu^{0}_{0}<28.2$ 
leaves 43 galaxies, to which we have added five galaxies from 
\citet{Sakai:etal:04}. 
While \citeauthor{Karachentsev:etal:04} have adopted a uniform
zero-point of the TRGB of $M_{B}=-4.05\mag$, \citeauthor{Sakai:etal:04}
have applied small corrections for metallicity, but their mean
zero-point of $-4.01\mag$ is sufficiently close not to make a
difference in the following. 
One may wonder whether the zero-point of the TRGB method is indeed
stable over a magnitude range of the galaxies from $-10$ to $-20\mag$,
i.e.\ a factor of $10^{4}$ in luminosity. Tests show that $\langle
H_{0}\rangle$ varies by merely $\sim\pm3\%$ if only the faintest or
only the brightest galaxies (giving a lower $H_{0}$) are considered. 
If real, this may be a metallicity or a population size effect, but it
is small enough to be neglected here. 

\begin{figure*}[p] 
     \epsscale{1.0}
     \plotone{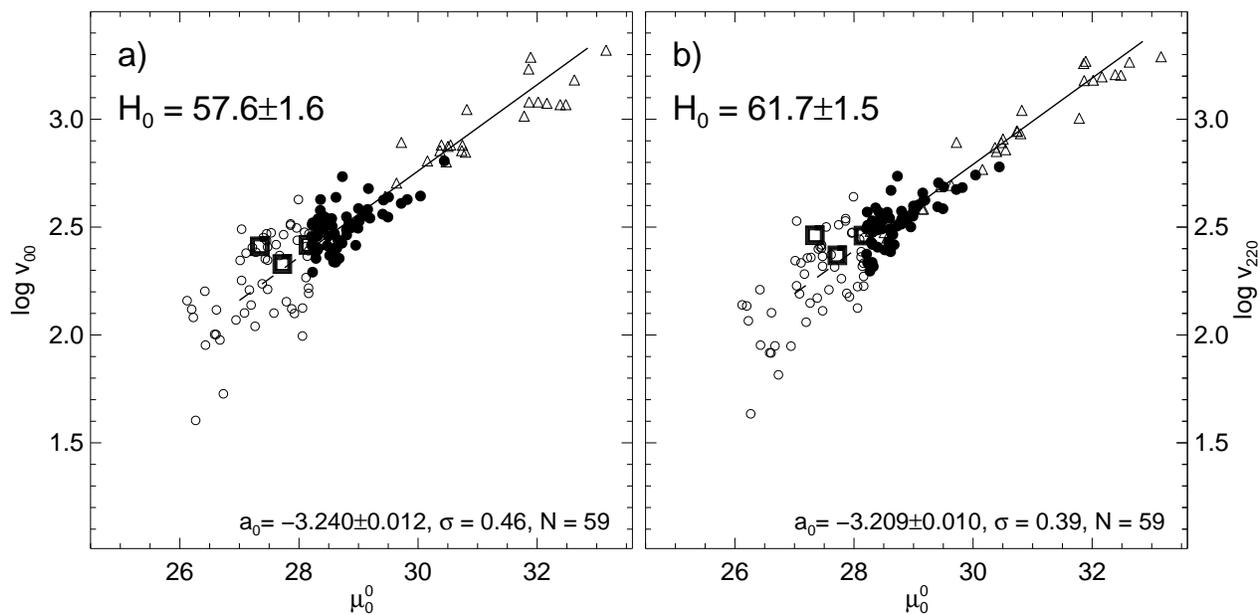} 
     \caption{The distance-calibrated Hubble diagram of galaxies with
     TRGB distance moduli (referred to the barycenter of the Local
     Group) and $\log v_{00}$ (also referred to the barycenter). Open
     and closed dots are for $\mu^{0}_{0}\lessgtr28.2$. The adopted
     SNe\,Ia and Cepheids of Fig.~\ref{fig:06} outside $\alpha_{\rm
     M87}=25^{\circ}$ are shown as open triangles ($\triangle$) for
     comparison. The squares stand for the M\,81, Cen\,A, and IC\,342
     groups. The full line is a fit to the TRGB galaxies with
     $(m-M)^{0}>28.2$.  b) Same as a), but corrected for
     Virgocentric infall.} 
\label{fig:TRGBkara}
\end{figure*}

     The 59 galaxies (excluding the deviating case D634-03) 
with TRGB distances $\mu^{0}_{0}>28.2$ are plotted in a
distance-calibrated Hubble diagram in Figure~\ref{fig:TRGBkara}
(filled symbols).  Using $v_{00}$ velocities as seen from the Local
Group barycenter 
yields $H_{0}=57.6\pm1.6$, which is increased to the adopted value of
$H_{0}=61.7\pm1.5$ after correction for Virgocentric infall. 
The dispersion about the Hubble line of $\sigma_{m}=0.39\mag$, which
must be caused mainly by peculiar motions, corresponds to 
$\partial v/v = 0.20$ or $v_{\rm pec}=125\kms$ at 10\,Mpc and 
$v_{\rm pec}=55\kms$ at 4.4\,Mpc.  
\clearpage
\clearpage

     It may appear puzzling that \citet{Karachentsev:05} derived
$H_{0}=72$ from six very nearby groups with distances from the TRGB
and other distance indicators, and $H_{0}=68$
\citep{Karachentsev:etal:05} from 110 {\em field\/} galaxies with
distances $2.5\la r\la7\;$Mpc based on TRGB {\em and\/} old Cepheid
distances. If, however, we restrict the sample to the 59 field
galaxies from above with TRGB distances larger than $4.4\;$Mpc and
proceed along the precepts of \citet{Karachentsev:etal:05}, i.e.\ with
distances as seen from the Galaxy (not the barycenter of the Local
Group) and velocities reduced to the barycenter following
\citet{Karachentsev:Makarov:01}, we obtain $H_{0}=58.2\pm2.4$, which
must be compared with $H_{0}=57.6\pm1.6$ from above. We can therefore
not reproduce values of $H_{0}=72$ or 68 using {\em only\/} TRGB
distances. 

     In the very nearby distance range of $27.0<\mu^{0}_{0}<28.2$,
i.e.\ between 2.5 and 4.4\,Mpc, are 89 galaxies with TRGB distances
from \citet{Karachentsev:etal:04,Karachentsev:etal:05}. Of these 45
galaxies are assigned to the M\,81, Cen\,A, and IC\,342 groups as
unquestionable members by \citet{Karachentsev:05}.
Relevant mean parameters of the three groups are compiled in
Table~\ref{tab:08} and their position in the Hubble diagram is shown
in Figure~\ref{fig:TRGBkara} (squares).
Column~(2) of Table~\ref{tab:08} gives the number of available
TRGB distances, whose mean is shown in column~(3). The velocity 
$\langle v_{\odot}\rangle$ in column~(5) is the mean over all group
members with known velocity. The individual values of the Hubble
parameter $H_{i}$ are in column~(9). The distance in km\,s$^{-1}$ of the
mean group velocity $\langle v_{220}\rangle$ from the Hubble line with
$H_{0}=60.3$ is in column~(10). 
While the first two groups in Table~\ref{tab:08} deviate by less than
$30\kms$ from the Hubble line, the IC\,342 group has an excess
velocity of $113\kms$, which may support the view that the galaxy was
early-on ejected from the Local Group \citep{Byrd:etal:94}.

     The remaining 44 field galaxies in the nearest distance interval
($27.0<\mu^{0}_{0}<28.2$), shown as open dots in
Fig.~\ref{fig:TRGBkara}, give 
$H_{0}=63.8\pm3.5$ from $v_{00}$ and $H_{0}=64.0\pm3.0$ using
$v_{220}$ velocities. The dispersion is reduced in the latter case to
$\sigma=0.67\mag$, which translates into random velocities of $65\kms$
at a median distance of 3.2\,Mpc. The data are fully consistent with
the conclusion that the underlying Hubble flow is linear down to
scales of $\sim\!2.5$\,Mpc. At still smaller distances the Hubble line
curves downwards as a clear effect of the gravitational pull of the
Local Group \citep{Lynden-Bell:81,Sandage:86}.

     Following \citet{Byrd:etal:94}
\citet{Chernin:etal:04,Chernin:etal:05} have performed model
calculations of a possibly violent formation process of the Local
Group whereby (some of) the dwarf galaxies may have been ejected into
the nearby field. They are predicted to have Hubble ratios of
$\sim\!90$ and small velocity dispersion. At most a handful of dwarfs
in Figure~\ref{fig:TRGBkara} (open symbols) match this prediction.

     The near agreement to better than 5\% between $H_{0}$ from local
Cepheids and SNe\,Ia ($H_{0}=60.9\pm1.3$) as well as from distant
SNe\,Ia ($H_{0}=62.3\pm1.3$) on the one hand and $H_{0}$ from the {\em
  independent\/} TRGB ($H_{0}=61.7\pm1.5$) on the other hand must be
emphasized. It suggests, in fact, that the {\em combined\/} systematic
error of the Cepheids and of the TRGB is not larger than
$\sim\!0.10\mag$. In \S~\ref{sec:H0cosmic:error} the systematic error of
$H_{0}$ has been estimated to be $0.17\mag$. This appears now like a
generous upper limit.

     The constancy of $H_{0}$ over scales from $\sim\!2.5$ to
200\,Mpc, with significant deviations only in and around bound
structures, has been a puzzle for a long time
\citep{Sandage:etal:72,Sandage:86,Sandage:99}. The solution that the
vacuum energy may be the dominant effect is further discussed in
\S~\ref{sec:conclusions}.

\section{PHYSICAL DISTANCE DETERMINATIONS}
\label{sec:physical}
Most astronomical distance determinations need some known distance as
a reference point. In this sense even distances from trigonometric
parallaxes require  knowledge of the Astronomical Unit. In contrast,
there are objects whose distance can be determined 
from only their physical (or geometrical) properties without reference
to any astronomical distance. These are referred here to as ``physical''
distances. 

     A typical example  are the moving-atmosphere
distances of Cepheids (the BBW method) which contribute to the
zero-point definition of the Galactic $P$-$L$ relation. Another
example is the water-maser distance of NGC\,4258
\citep{Herrnstein:etal:99} as discussed in
\citeauthor{Saha:etal:06}. As more of them become available they too
will contribute to the zero-point definition of the $P$-$L$ relation
of Cepheids. The eclipsing binary distance of M\,31 of 
$(m-M)^{0}=24.44\pm0.12$ \citep{Ribas:etal:05} is in satisfactory
agreement with its Cepheid distance of $24.54$ in
\citeauthor{Saha:etal:06}. 

     Of particular interest for the present paper are SN\,Ia models
which predict the luminosity of SNe\,Ia at maximum. Typical results
are $M_{B}\approx M_{V}=-19.50$ (\citealt{Branch:98} for a
review) which agrees spectacularly well with our empirically
determined value of $M_{V}=-19.46$ in \S~\ref{sec:Mabs:adopted}.   

     ``Expanding-atmosphere parallaxes'' (EPM) of SNe\,II have been
determined by several authors. The difficulty here is that the
radiation transport in fast moving atmospheres poses enormous
problems. 
Five EPM distances of \citet{Schmidt:etal:94} for the five galaxies
for which also Cepheid distances are available
(\citeauthor{Saha:etal:06}, Table~A1) show good agreement to
$0.09\pm0.13$ (the EPM distances being formally larger), 
but the larger EPM distances suggest an unrealistic increase of
$H_{0}$ with distance. Only two of Hamuy's (\citeyear{Hamuy:01})
SN\,II distances can be compared with Cepheids; the Cepheid distances
are larger by $\sim\!0.5\mag$. 
A recent progress in the ``Spectral-fitting expanding
atmosphere method'' (SEAM) of the type II SN\,1999em
\citep{Baron:etal:04} gives $(m-M)^{0}=30.48\pm0.30$, which compares
favorably with the Cepheid distance of its parent galaxy NGC\,1637 of
$\mu^{0}_{Z}=30.40$. -- A different method has been developed for type
IIP SNe by \citet{Nadyozhin:03}. It uses the expansion velocity and
duration of the plateau phase as well as the tail magnitude to yield
for eight SNe\,IIP a value of $H_{0}=55\pm5$. However, the result
depends on the assumption that the $^{56}$Ni energy equals the
expansion energy, which requires further testing.

     Much effort has gone into the determination of $H_{0}$ from the
Sunyaev-Zeldovich (SZ) effect. A typical result is $H_{0}=60\pm3$, yet
with a systematic error of $\pm18$ \citep{Carlstrom:etal:02}. Also
more recent investigations give similar values and errors
\citep[e.g..][]{Udomprasert:etal:04,Jones:etal:05}. 

     Gravitationally lensed quasars with two or more images provide
distances if the redshift of the lens and of the quasar and the delay
time between the images are known. Unfortunately the solution is
degenerate as to the distance {\em and\/} the mass distribution of the
lens. Values of $H_{0}$ for different sources and by different
authors therefore vary still between $48<H_{0}<75$
\citep[e.g.][]{Saha:03,Koopmans:etal:03,Kochanek:Schechter:04,York:etal:05,Magain:05}.   
\citet{SahaP:etal:06} 
state that the current data are consistent with $H_0$ anywhere
between $60$ and $80$ at the $1\sigma$ level depending on the mass
model of the lens.
     
     The last-mentioned method does not in principle provide the
(present) value of $H_{0}$, but $H_{z}$ at the epoch of the quasar. To
obtain $H_{0}$ some assumptions on $\Omega_{\rm M}$ and
$\Omega_{\Lambda}$ are necessary. The disadvantage of long look-back
times for the determination of $H_{0}$ becomes most pronounced in case
of the Fourier spectrum of the CMB acoustic waves, where a number of
free parameters must simultaneously be solved for. 
The solution for $H_{0}$ depends
therefore on the number of free parameters allowed for and on some
priors forced on the data as well as on the observations used.
A six-parameter solution of the WMAP data with some priors and
additional observational constraints has yielded $H_{0}=73\pm3$
\citep{Spergel:etal:03,Spergel:etal:06}. 
This has often been taken as a confirmation
of $H_{0}=72\pm8$ as obtained from various distance indicators by
\citet{Freedman:etal:01}, and has led to the opinion that the problem
of $H_{0}$ has been solved. We disagree. The actual situation has been
illustrated by \citet{Rebolo:etal:04} who have used the Very
Small Array and WMAP data to derive $H_{0}=66\pm7$ allowing for twelve
free parameters and no priors. Clearly a strong motive to further
reduce the systematic error of $H_{0}$ by conventional means comes
from the desire to use the Hubble constant itself as a reliable prior
for the interpretation of the CMB spectrum.  

     A value of $H_{0}=62.3$ corresponds in an $\Omega_{\rm M}=0.3$,
$\Omega_{\Lambda}=0.7$ universe to an expansion age of $15.1\;$Gyr,
which may be compared with the age of M\,92 of $13.5\;$Gyr
\citep{VandenBerg:etal:02} and the Th/Eu age of the Galactic halo of
$\sim\!15\;$Gyr \citep{Pagel:01}. Ultra-metal-poor giants yield
radioactive ages between $14.2\pm3.0$ to $15.6\pm4.0\;$Gyr
\citep{Cowan:etal:99,Westin:etal:00,Truran:etal:01,Sneden:etal:03}. A
high-weight determination of the U/Th age of the Milky Way gives
$14.5\pm2.5\;$Gyr \citep{Dauphas:05}. All these values must, of
course, still to be increased by the gestation time of the chemical
elements.

\section{CONCLUSIONS}
\label{sec:conclusions}
     (1) The final result of our HST collaboration, ranging over 15
years, is that 
\begin{equation}
   H_{0}(\mbox{cosmic}) = 62.3 \pm 1.3 \;(\mbox{random}) \pm 5.0 
      \;(\mbox{systematic})
\label{eq:08}
\end{equation}
based on 62 SNe\,Ia with $3000 < v_{\rm CMB} < 20\,000\kms$ and on 10
luminosity-calibrated SNe\,Ia. All SNe\,Ia have been corrected for
Galactic and internal absorption and are normalized  to decline rate
$\Delta m_{15}$ and color (\citeauthor{Reindl:etal:05}).  
The weighted mean luminosities of the 10 calibrators of
$M_{B}=-19.49$, $M_{V}=-19.46$, and $M_{I}=-19.22$
(Table~\ref{tab:04}) are based on metallicity-corrected Cepheid
distances (\citeauthor{Saha:etal:06}) from the new $P$-$L$ relations
of the Galaxy and LMC (Paper~I \& II).   

     (2) The local value of $H_{0}$ ($300\la v_{220} < 2000\kms$) is
\begin{equation}
   H_{0}(\mbox{local}) = 60.9 \pm 1.3\; (\mbox{random}) \pm 5.0 
      \;(\mbox{systematic})
\label{eq:09}
\end{equation}
from 25 Cepheid and 16 SNe\,Ia distances, involving a total of 34
different galaxies. Their distances are related to the
barycenter of the Local Group and their observed velocities are
corrected for a self-consistent Virgocentric infall model with a local
infall vector of $220\kms$.   
The local value of $H_{0}$ is supported by the mean distances and mean
velocities $\langle v_{220}\rangle$ of the Virgo and Fornax cluster
(Table~\ref{tab:07}).   

     (3) The values of $H_{0}$ under 1) and 2) find strong support by
TRGB distances which constitute an {\em independent\/} Pop.~II
distance scale. Forty-seven TRGB distances in the range from 4.4\,Mpc
to 10\,Mpc yield $H_{0}=60.3\pm1.8$. This may suggest that the
systematic error in equations~(\ref{eq:08}) and (\ref{eq:09}) has been
overestimated. 

     (4) The constancy of $H_{0}$ from global cosmic scales down to
4.4\,Mpc or even 2.5\,Mpc \citep[see also][]{Ekholm:etal:01} in spite
of the inhomogeneous mass distribution requires a special agent. 
Vacuum energy as the solution has been proposed by several authors
\citep[e.g.][]{Baryshev:etal:01,Chernin:01,Chernin:etal:03a,Chernin:etal:03b,Thim:etal:03,Teerikorpi:etal:05}.
No viable alternative to vacuum energy is known at present. The
quietness of the Hubble flow lends support for the existence of vacuum
energy.

     (5) The modulating effect of bound structures and their
surroundings on the Hubble flow is seen in the immediate neighborhood
of the Local Group and particularly clearly in the successful
Virgocentric flow model (see Fig.~\ref{fig:04} and \ref{fig:05}).

     (6) Random velocities of field galaxies at large distances
($5000<v_{\rm cmb}<20\,000\kms$) are confined by 20 SNe\,Ia in E/S0
parent galaxies (and hence with little internal absorption). 
Their scatter about the Hubble line is $\sigma_{I}=0.10\mag$
(\citeauthor{Reindl:etal:05}) which implies $\partial v/v \le 0.05$
or $250\kms$ at a distance of $5000\kms$. This is a strict upper limit
because no allowance for the intrinsic and observational scatter of
the normalized SN\,Ia luminosities has been made. At intermediate
distances ($300<v_{220}<2000\kms$) the scatter from Cepheids and
SNe\,Ia is $0.33\mag$ without any allowance for observational
errors. Therefore $\partial v/v \le 0.16$, which corresponds to
peculiar velocities of $\le160\,(80)\kms$ at a distance of
$1000\,(500)\kms$. TRGB distances give $v_{\rm pec}=60\kms$ at a
distance of $300\kms$. In the still closer neighborhood of the Local
Group the velocity dispersion is similar, emphasizing again the
quietness of the Hubble flow. 
 
     (7) The adopted value of $H_{0}=62.3$ rests about equally on two
zero-points. (i) The zero-point of the Galactic $P$-$L$ relation of
Cepheids which is determined with equal weights from 33 purely
physical moving-atmosphere (BBW) parallaxes and 36 Cepheids in
Galactic clusters, which are fitted to the ZAMS of the Pleiades at 
$(m-M)^{0}=5.61$. And (ii) The zero-point of the LMC $P$-$L$ relation
which is based on an adopted LMC modulus of 18.54 from various
determinations, but excluding methods which involve the $P$-$L$
relation itself. 

     (8) The adopted metallicity corrections of the Cepheid distances
are supported by model calculations by \citet{Fiorentino:etal:02} and 
\citet{Marconi:etal:05} who show that the instability strip shifts
redwards in the $\log L - \log T_{\rm e}$ plane (at $L=\mbox{const.}$) 
as the metallicity increases similar to earlier models by authors
cited in \S~\ref{sec:intro}, but now also for a wide range of $\Delta
Y/ \Delta Z$. Moreover, the metallicity corrections are supported by
comparing the metallicity-corrected moduli $\mu^{0}_{Z}$ with
independent TRGB and velocity distances. Additional positive tests
are provided by the metal-rich and metal-poor Cepheids in NGC\,5457
(M\,101) and by the (near) independence of the SN\,Ia luminosities on
the metallicity of their parent galaxies.

     (9) Several previous authors, listed in Table~\ref{tab:06}, have
found from SNe\,Ia values of $H_{0}$ in statistical agreement with
equation~(\ref{eq:08}). The present determination, however, is based
on larger data sets and, as we believe, on more realistic $P$-$L$
relations of Cepheids. Significantly larger values of $H_{0}$ in 
Table~\ref{tab:06} are mainly due to the adopted (small) Cepheids
distances by others -- and not due to the specific treatment of the
data on SNe\,Ia per se, such as the absorption corrections in the 
parent galaxies, or the normalizations to decline rate and color. 
Already \citet{Germany:etal:04} have pointed out the large spread of
SNe\,Ia-based determinations of $H_{0}$ in the literature is almost
entirely due to systematic errors of the Cepheid distances.
Had \citet{Freedman:etal:01}, e.g., used our Cepheid distances for
their six calibrating SNe\,Ia and their version of the SN\,Ia
Hubble diagram, they would have found $H_{0}=60.5$ instead of 72. 

     (10) The value of $H_{0}(\mbox{cosmic}$) corresponds in a
standard $\Lambda\,$CDM ($\Omega_{\rm M}=0.3,\Omega_{\Lambda}=0.7$) 
model to an expansion age of $15.1\;$Gyr,
giving a sufficient time frame for even the oldest Galactic globular
clusters and highest radioactive ages.

\acknowledgments

Allan~S. needs to make a statement:
``Much of the work in this paper has been done by G.A.T., and,
although input has been made by all the signed authors, under normal
circumstances Tammann would be the lead author. However, G.A.T. and
Abhijit~S. have conspired to offer as an 80th year birthday gift to me
the first-author-place for which I have acquiesced to emphasize the 43
years of a most enjoyable collaboration with G.A.T. and the almost 20
years with Abhijit on this problem.
Ours has been a most unusual, productive, and magic scientific time
together, working within the SNe\,Ia HST collaboration.''
Allan S.'s collaborators thank him for having guided this project as a
PI over most of its duration and for his inspiration and incessant
drive. 
We thank collectively the many individuals at STScI who over time
made the original observations with HST possible.
We also thank several collaborators, particularly Lukas Labhardt and
Frank Thim, who have joined us at different stages of our long
endeavor.
It is a pleasure to thank Susan Simkin, scientific editor, for her
sensitivity in seeing this paper on the continuing highly charged
debate over $H_{0}$ through the editorial process. We also thank the
two referees whom she chose for their neutrality, one anonymous and
the other Nicholas Suntzeff for their several suggestions for clarity
which we have followed.
Abhijit~S. thanks for support provided by NASA of the
retro-active analysis WFPC2 photometry zero-points through grant
HST-AR-09216.01A from the Space Telescope Science Institute, which is
operated by the Association of Universities for Research in Astronomy,
Inc., under NASA contract  NAS 5-26555.  
Bernd~R. thanks the Swiss National Science Foundation for financial
support.
Allan~S. thanks the Carnegie Institution for post-retirement
facilities support.

\clearpage


\clearpage

\begin{center}
\begin{deluxetable}{llrrccccccccc}
\rotate
\tablewidth{0pt}
\tabletypesize{\scriptsize}
\tablecaption{Parameters of Galaxies with SNe\,Ia and Cepheid
  Distances.\label{tab:01}}  
\tablehead{
 \colhead{SN} & 
 \colhead{Galaxy} & 
 \colhead{$v_{0}$} &
 \colhead{$v_{220}$} &
 \colhead{[O/H]$_{\rm old}$} &
 \colhead{[O/H]$_{\rm Sakai}$} &
 \colhead{N} &
 \colhead{$\langle\log P\rangle$} &
 \colhead{$\mu^{0}(\mbox{Gal})$} &
 \colhead{$\mu^{0}(\mbox{LMC})$} &
 \colhead{$\mu^{0}_{Z}(\mbox{M/F})$} &
 \colhead{$\mu^{0}_{Z}$} &
 \colhead{$\epsilon(\mu^{0}_{Z})$} \\
 \colhead{(1)}  & \colhead{(2)}  &
 \colhead{(3)}  & \colhead{(4)}  &
 \colhead{(5)}  & \colhead{(6)}  &
 \colhead{(7)}  & \colhead{(8)}  &
 \colhead{(9)}  & \colhead{(10)} &
 \colhead{(11)} & \colhead{(12)} &
\colhead{(13)} 
} 
\startdata
1937C   & IC\,4182     & 342    & 303    & 8.40 & 8.20 & 13 & 1.387 & 28.51 & 28.32 & 28.25 & 28.21 & 0.10    \\
1960F   & NGC\,4496A   & 1573   & 1152   & 8.77 & 8.53 & 39 & 1.514 & 31.24 & 30.99 & 31.17 & 31.18 & 0.10    \\
1972E   & NGC\,5253    & 156    & 170    & 8.15 & 8.15 & 5  & 1.029 & 28.11 & 28.09 & 27.89 & 28.05 & 0.27    \\
1974G   & NGC\,4414    & 691    & 1137   & 9.20 & 8.77 & 10 & 1.526 & 31.55 & 31.29 & 31.63 & 31.65 & 0.17    \\
1981B   & NGC\,4536    & 1645   & (1152) & 8.85 & 8.58 & 27 & 1.566 & 31.26 & 30.98 & 31.20 & 31.24 & 0.10    \\
1989B   & NGC\,3627    & 597    & 427    & 9.25 & 8.80 & 22 & 1.452 & 30.41 & 30.19 & 30.53 & 30.50 & 0.10    \\
1990N   & NGC\,4639    & 901    & 1152   & 9.00 & 8.67 & 12 & 1.552 & 32.13 & 31.86 & 32.13 & 32.20 & 0.10    \\
1991T   & NGC\,4527    & 1575   & (1152) & 8.75 & 8.52 & 19 & 1.498 & 30.84 & 30.59 & 30.76 & 30.76 & 0.20    \\
1994ae  & NGC\,3370    & 1169   & 1611   & 8.80 & 8.55 & 64 & 1.548 & 32.42 & 32.14 & 32.35 & 32.37 & 0.10    \\
1998aq  & NGC\,3982    & 1202   & 1510   & 8.75 & 8.52 & 15 & 1.502 & 31.94 & 31.69 & 31.86 & 31.87 & 0.15    \\
1998bu  & NGC\,3368    & 760    & 708    & 9.20 & 8.77 & 7  & 1.467 & 30.25 & 30.02 & 30.34 & 30.34 & 0.11    \\
1999by  & NGC\,2841    & 716    & 895    & 8.80 & 8.55 & 18 & 1.445 & 30.79 & 30.57 & 30.74 & 30.75 & 0.10    \\ 
\enddata
\end{deluxetable}
\end{center}

\clearpage

\def\baselinestretch{1.1}
\begin{deluxetable}{llcrllrrr}
\tablewidth{0pt}
\tabletypesize{\scriptsize}
\tablecaption{Parameters of SNe\,Ia with Cepheid Distances.\label{tab:02}} 
\tablehead{
 \colhead{SN} & 
 \colhead{$\Delta m_{15}$} & 
 \colhead{$E_{\rm Gal}$} &
 \colhead{$E_{\rm host}$} &
 \colhead{$(B\!-\!V)$} &
 \colhead{$(V\!-\!I)$} &
 \colhead{$m_{B}^{\rm corr}$} &
 \colhead{$m_{V}^{\rm corr}$} &
 \colhead{$m_{I}^{\rm corr}$} \\
 \colhead{(1)}  & \colhead{(2)}  &
 \colhead{(3)}  & \colhead{(4)}  &
 \colhead{(5)}  & \colhead{(6)}  &
 \colhead{(7)}  & \colhead{(8)}  &
 \colhead{(9)}  
} 
\startdata
1937C   & 0.85   & 0.014 & $-$0.022 & $-$0.012 & \nodatl  &  8.97\,(09) &  8.99\,(11) & \nodatl      \\ 
1960F   & 0.87   & 0.025 &    0.099 & $-$0.034 & \nodatl  & 11.28\,(15) & 11.31\,(20) & \nodatl      \\ 
1972E   & 1.05   & 0.056 & $-$0.050 & $-$0.006 & $-$0.316 &  8.46\,(14) &  8.49\,(15) &  8.77\,(19)  \\ 
1974G   & 1.11   & 0.019 &    0.161 & $+$0.000 & \nodatl  & 11.79\,(05) & 11.82\,(05) & \nodatl      \\ 
1981B   & 1.13   & 0.018 &    0.037 & $+$0.005 & \nodatl  & 11.79\,(05) & 11.82\,(05) & \nodatl      \\ 
1989B   & 1.31   & 0.032 &    0.311 & $+$0.007 & $-$0.162 & 10.93\,(05) & 10.95\,(05) & 11.11\,(05)  \\ 
1990N   & 1.05   & 0.026 &    0.034 & $-$0.040 & $-$0.298 & 12.56\,(05) & 12.59\,(05) & 12.89\,(05)  \\ 
1991T   & 0.94   & 0.022 &    0.199 & $-$0.031 & $-$0.411 & 10.98\,(05) & 11.00\,(05) & 11.39\,(05)  \\ 
1994ae$^*$  & 0.90   & 0.030 &    0.034 & $-$0.064 & $-$0.294 & 12.98\,(05) & 13.01\,(05) & 13.31\,(05)  \\ 
1998aq$^*$  & 1.05   & 0.014 & $-$0.048 & $-$0.066 & $-$0.214 & 12.54\,(05) & 12.56\,(05) & 12.82\,(05)  \\ 
1998bu  & 1.15   & 0.025 &    0.279 & $+$0.056 & $-$0.191 & 11.01\,(05) & 11.04\,(05) & 11.14\,(05)  \\
1999by  & 1.90   & 0.016 & \nodatl  & $+$0.494 & $+$0.219 & 13.59\,(05) & 13.10\,(05) & 12.88\,(05)  \\
\enddata
\tablecomments{$^*$ The photometry is from \citet{Riess:etal:05} and
  reduced as in \citeauthor{Reindl:etal:05}.}                          
\end{deluxetable}

\setlength\textheight{9.0in}%
\begin{center}
\begin{deluxetable}{lccc|ccc|ccc|ccc}
\rotate
\tablewidth{0pt}
\tabletypesize{\scriptsize}
\tablecaption{Absolute Magnitudes of SNe\,Ia and Solutions for 
              $H_{0}$.\label{tab:03}} 
\tablehead{
 & & & & & & & & & & & & \\[-15pt]
 & & & & & & & & & & & & \\[-8pt]
& \multicolumn{3}{c|}{from Gal. $P$-$L$} &
  \multicolumn{3}{c|}{from LMC $P$-$L$}  &
  \multicolumn{3}{c|}{from M/F $P$-$L$}   &
  \multicolumn{3}{c}{from $\mu^{0}_{Z}$} \\
 {SN} & 
 {$M_{B}$} & 
 {$M_{V}$} & 
 {$M_{I}$} & 
 {$M_{B}$} & 
 {$M_{V}$} & 
 {$M_{I}$} & 
 {$M_{B}$} & 
 {$M_{V}$} & 
 {$M_{I}$} & 
 {$M_{B}$} & 
 {$M_{V}$} & 
 {$M_{I}$} \\
 {(1)}  & {(2)}  &
 {(3)}  & {(4)}  &
 {(5)}  & {(6)}  &
 {(7)}  & {(8)}  &
 {(9)}  & {(10)} &
 {(11)} & {(12)} &
 {(13)}
}
\startdata
 & & & & & & & & & & & & \\[-16pt]
 & & & & & & & & & & & & \\[-8pt]
1937C    & -19.54\,(13) & -19.52\,(15) & \nodata      & -19.35\,(13) & -19.33\,(15) & \nodata      & -19.28\,(13) & -19.26\,(15) & \nodata      & -19.24\,(13) & -19.22\,(15) & \nodata       \\        
1960F    & -19.96\,(18) & -19.93\,(22) & \nodata      & -19.71\,(18) & -19.68\,(22) & \nodata      & -19.89\,(18) & -19.86\,(22) & \nodata      & -19.90\,(18) & -19.87\,(22) & \nodata       \\ 
1972E    & -19.65\,(30) & -19.62\,(31) & -19.34\,(33) & -19.63\,(30) & -19.60\,(31) & -19.32\,(33) & -19.43\,(30) & -19.40\,(31) & -19.12\,(33) & -19.59\,(30) & -19.56\,(31) & -19.28\,(33)  \\
1974G    & -19.76\,(18) & -19.73\,(18) & \nodata      & -19.50\,(18) & -19.47\,(18) & \nodata      & -19.84\,(18) & -19.81\,(18) & \nodata      & -19.86\,(18) & -19.83\,(18) & \nodata       \\ 
1981B    & -19.47\,(11) & -19.44\,(11) & \nodata      & -19.19\,(11) & -19.16\,(11) & \nodata      & -19.41\,(11) & -19.38\,(11) & \nodata      & -19.45\,(11) & -19.42\,(11) & \nodata       \\ 
1989B    & -19.48\,(11) & -19.46\,(11) & -19.30\,(11) & -19.26\,(11) & -19.24\,(11) & -19.08\,(11) & -19.60\,(11) & -19.58\,(11) & -19.42\,(11) & -19.57\,(11) & -19.55\,(11) & -19.39\,(11)  \\
1990N    & -19.57\,(11) & -19.54\,(11) & -19.24\,(11) & -19.30\,(11) & -19.27\,(11) & -18.97\,(11) & -19.57\,(11) & -19.54\,(11) & -19.24\,(11) & -19.64\,(11) & -19.61\,(11) & -19.31\,(11)  \\
1994ae   & -19.44\,(11) & -19.41\,(11) & -19.11\,(11) & -19.16\,(11) & -19.13\,(01) & -18.83\,(11) & -19.37\,(11) & -19.34\,(11) & -19.04\,(01) & -19.39\,(11) & -19.36\,(11) & -19.06\,(11)  \\
1998aq   & -19.40\,(16) & -19.38\,(16) & -19.12\,(16) & -19.15\,(16) & -19.13\,(16) & -18.87\,(16) & -19.32\,(16) & -19.30\,(16) & -19.04\,(16) & -19.33\,(16) & -19.31\,(16) & -19.05\,(16)  \\
1998bu   & -19.24\,(12) & -19.21\,(12) & -19.11\,(12) & -19.01\,(12) &
-18.98\,(12) & -18.88\,(12) & -19.33\,(12) & -19.30\,(12) &
-19.20\,(12) & -19.33\,(12) & -19.30\,(12) & -19.20\,(12)  \\[2pt]
\tableline
 & & & & & & & & & & & & \\[-8pt]
mean     & -19.55\,(06) & -19.52\,(06) & -19.20\,(04) & -19.33\,(07) & -19.30\,(07) & -18.99\,(08) & -19.50\,(07) & -19.48\,(07) & -19.18\,(06) & -19.53\,(07) & -19.50\,(07) & -19.22\,(06)  \\
weighted & -19.50\,(04) & -19.47\,(04) & -19.19\,(05) & -19.26\,(04) & -19.22\,(04) & -18.94\,(05) & -19.48\,(04) & -19.45\,(04) & -19.20\,(05) & -19.49\,(04) & -19.46\,(04) & -19.22\,(05)  \\[2pt]
\tableline
 & & & & & & & & & & & & \\[-8pt]
$H_{0}$(mean)     & 60.7\,(1.7) & 60.8\,(1.7) & 62.7\,(1.2) & 67.1\,(2.2) & 67.3\,(2.2) & 69.0\,(2.6) & 61.8\,(2.0) & 61.9\,(2.0) & 63.2\,(1.8) & 61.2\,(2.0) & 61.4\,(2.0) & 62.1\,(1.4) \\
$H_{0}$(weighted) & 62.1\,(1.2) & 62.2\,(1.2) & 63.0\,(1.5) & 69.3\,(1.3) & 69.8\,(1.3) & 70.6\,(1.6) & 62.7\,(1.2) & 62.8\,(1.2) & 62.7\,(1.5) & 62.4\,(1.2) & 62.5\,(1.2) & 62.1\,(1.2) \\[2pt]
\tableline
\noalign{\smallskip}
\noalign{\smallskip}
\multicolumn{13}{c}{\footnotesize{non-standard spectra}}\\
\noalign{\smallskip}
\tableline
 & & & & & & & & & & & & \\[-8pt]
1991T    & -19.86\,(21) & -19.84\,(21) & -19.45\,(21) & -19.61\,(21) & -19.59\,(21) & -19.20\,(21) & -19.78\,(21) & -19.76\,(21) & -19.37\,(21) & -19.78\,(21) & -19.76\,(21) & -19.37\,(21)  \\ 
1999by   & -17.20\,(11) & -17.69\,(11) & -17.91\,(11) & -16.98\,(11) & -17.47\,(11) & -17.69\,(11) & -17.15\,(11) & -17.64\,(11) & -17.86\,(11) & -17.16\,(11) & -17.65\,(11) & -17.87\,(11)  \\[-8pt]
 & & & & & & & & & & & & \\[-2.5pt]
\enddata                          
\tablecomments{$H_{0}$ from $c_{B}=0.693\pm0.004$,
  $c_{V}=0.688\pm0.004$, $c_{I}=0.637\pm0.004$ (from
  \citeauthor{Reindl:etal:05}) and $\log H_{0}=0.2\cdot M_{\lambda} +
  c_{\lambda} + 5$}                           
\end{deluxetable}
\end{center}

\clearpage
\setlength\textheight{8.4in}%

\begin{deluxetable}{lccc}
\tablewidth{0pt}
\tabletypesize{\footnotesize}
\tablecaption{Weighted, Metallicity-Corrected Mean Absolute
  Magnitudes of SNe\,Ia. The Error of the Mean is $0.04\mag$ for All
  Entries.\label{tab:04}}  
\tablehead{
  & 
 \colhead{$M_{B}$} & 
 \colhead{$M_{V}$} & 
 \colhead{$M_{I}$} 
} 
\startdata
1) from $\mu^{0}$(Gal)                            & $-19.50$ & $-19.47$ & $-19.19$ \\
2) from $\mu^{0}_{Z}$(LMC)                        & $-19.36$ & $-19.32$ & $-19.07$ \\
3) from $\mu^{0}_{Z}$(M/F)                        & $-19.48$ & $-19.45$ & $-19.20$ \\
{\bf 4) from \boldmath{$\mu^{0}_{Z}$} (\citeauthor{Saha:etal:06})} &  \boldmath{$-19.49$} & \boldmath{$-19.46$} & \boldmath{$-19.22$} \\
\noalign{\smallskip}
\tableline
\noalign{\smallskip}
mean of 1) $-$ 4)                                 & $-19.46$ & $-19.43$ & $-19.17$ \\
mean of 1), 3), \& 4)                             & $-19.49$ & $-19.46$ & $-19.20$ \\
\enddata
\end{deluxetable}

\clearpage

\begin{deluxetable}{rcccccc}
\tablewidth{0pt}
\tabletypesize{\footnotesize}
\tablecaption{Mean Properties of Calibrating and Distant
  SNe\,Ia.\label{tab:05}}   
\tablehead{
  & 
 \colhead{$N$} & 
 \colhead{$\langle(B\!-\!V)^{\rm corr}\rangle$} & 
 \colhead{$\langle(V\!-\!I)^{\rm corr}\rangle$} & 
 \colhead{$\langle E(B\!-\!V)_{\rm host}\rangle$} & 
 \colhead{$\langle\Delta m_{15}\rangle$} & 
 \colhead{$\langle$Host Galaxy Type$\rangle$} 
} 
\startdata
Calibrators     & 10 & $-0.024$ & $-0.235$ & $0.084$ & $1.05$ & $4.0$ \\ 
Distant SNe\,Ia & 62 & $-0.024$ & $-0.251$ & $0.057$ & $1.22$ & $1.5$ \\ 
\enddata
\end{deluxetable}

\clearpage

\begin{deluxetable}{lccc}
\tablewidth{0pt}
\tabletypesize{\footnotesize}
\tablecaption{Overview of $H_{0}$ Values from SNe\,Ia.\label{tab:06}}   
\tablehead{
 \colhead{authors} & 
 \colhead{cal} & 
 \colhead{dist} & 
 \colhead{$H_{0}$} \\
  & 
 \colhead{SNe} & 
 \colhead{SNe} & 
} 
\startdata
\citealt{Sandage:Tammann:82}    &  2 & 16 &   $50\pm7$  \\ 
\citealt{Capaccioli:etal:90}    & 10 &  5 &   $70\pm15$ \\ 
\citealt{Saha:etal:94}          &  1 & 34 &   $52\pm9$  \\ 
\citealt{Riess:etal:95}         &  1 & 13 &   $67\pm7$  \\
\citealt{Saha:etal:95}          &  3 & 34 &   $52\pm8$  \\ 
\citealt{Tammann:Sandage:95}    &  3 & 39 & $56.5\pm4$  \\ 
\citealt{Mould:etal:95}         &  6 & 21 &   $71\pm7$  \\ 
\citealt{Saha:etal:96}          &  4 & 39 & $56.5\pm3$  \\ 
\citealt{Hamuy:etal:96}         &  4 & 29 & $63.1\pm3.4\,\pm2.9$ \\ 
\citealt{Hoeflich:Khokhlov:96}  &  theory & 26 &   $67\pm9$ \\ 
\citealt{Saha:etal:97}          &  7 & 56 &   $58\pm8$ \\ 
\citealt{Saha:etal:99}          &  9 & 35 &   $60\pm2$ \\ 
\citealt{Tripp:Branch:99}       &  6/10 & 26/29 &   $62.9\pm4.7$ \\ 
\citealt{Suntzeff:etal:99}      &  8 & 40 & $63.9\pm2.2\,\pm3.5$ \\ 
\citealt{Phillips:etal:99}      &  6 & 40 & $63.3\pm2.2\,\pm3.5$ \\ 
\citealt{Jha:etal:99}           &  4 & 42 & $64.4\pm6.6\,\pm5.4$ \\
\citealt{Richtler:Drenkhahn:99} &  4 & 26 & $72\pm4$ \\
\citealt{Gibson:etal:00}        &  6 & 40 & $68\pm2\,\pm5$  \\
\citealt{Parodi:etal:00}        &  8 & 35 & $58.5\pm4.0$    \\
\citealt{Freedman:etal:01}      &  6 & 36 & $72\pm2\,\pm6$ \\
\citealt{Saha:etal:01}          &  9 & 35 & $58.7\pm2\,\pm6$ \\
\citealt{Altavilla:etal:04}     &  9 & 18-46 & $68-74$    \\
\citealt{Riess:etal:05}         &  4 & 68 & $73\pm4$      \\
\citealt{Wang:etal:06}          & 11 & 73 & $72\pm4$      \\
{\bf Present Paper}       & {\bf 10} & {\bf 62} & \boldmath{$62.3\pm1.3\,\pm5.0 $}
\enddata
\end{deluxetable}

\clearpage

\begin{deluxetable}{lllll}
\tablewidth{0pt}
\tabletypesize{\footnotesize}
\tablecaption{Distance of the Virgo and Fornax Cluster from
  Cepheids, SNe\,Ia, and 21cm-line Widths.\label{tab:07}}   
\tablehead{
  & 
 \multicolumn{2}{c}{Virgo} &
 \multicolumn{2}{c}{Fornax} \\
  &
 \colhead{Object} & 
 \colhead{$\mu_{\rm Z}^{0}$} &
 \colhead{Object} & 
 \colhead{$\mu_{\rm Z}^{0}$}
} 
\startdata
Cepheids         & N4321  & $31.18$ & N1326A & $31.17$ \\ 
                 & N4535  & $31.25$ & N1365  & $31.46$ \\
                 & N4548  & $30.99$ & N1425  & $31.96$ \\
                 & N4639  & $32.20$ &        &         \\
\noalign{\smallskip}
SNe\,Ia          & 1984A  & $31.15$ & 1980N  & $31.67$ \\
                 & 1990N  & $32.05$ & 1981D  & $31.30$ \\
                 & 1994D  & $31.30$ & 1992A  & $31.82$ \\
\multicolumn{2}{l}{21cm-line widths} & $31.58$ & & \\
\noalign{\smallskip}
$\langle\mu^{0}\rangle$       & & $31.47\pm0.16$ & & $31.56\pm0.13$ \\
$\langle\mu^{0}_{0}\rangle$   & & $31.51\pm0.16$ & & $31.56\pm0.13$ \\
$\langle v_{\odot}\rangle$  & & $1050\pm35^{1)}$         & & $1493\pm36^{2)}$ \\
$\langle v_{0}\rangle$        & & $932$         & & $1403$ \\
$\langle v_{00}\rangle$       & & $932$         & & $1403$ \\
$\langle v_{220}\rangle$      & & $1152\;(\pm35)$ & & $1371$ \\
$v_{\rm cosmic}$              & & $1175\pm30^{3)}$         & & \nodata \\
$H_{0}$                       & & $58.1\pm4.6$   & & $66.8\pm4.0$ \\
\enddata
\tablenotetext{1)}{\citet{Binggeli:etal:93}} 
\tablenotetext{2)}{\citet{Drinkwater:etal:01}} 
\tablenotetext{3)}{\citet{Jerjen:Tammann:93}. The value is inferred
  from the distance ratios between the Virgo cluster and more remote
  clusters whose CMB-corrected velocities are taken to define the
  cosmic expansion field.} 
\tablecomments{The Cepheid distances have been determined in
  \citeauthor{Saha:etal:06} (Table~A1) from the original Cepheid data by 
\citet[][for NGC\,4321]{Ferrarese:etal:97}, 
\citet[][for NGC\,4535]{Macri:etal:99}, 
\citet[][for NGC\,4548]{Graham:etal:99}, 
\citet[][for NGC\,4639]{Saha:etal:97}, 
\citet[][for NGC\,1326A]{Prosser:etal:99}, 
\citet[][for NGC\,1365]{Silbermann:etal:99},
and
\citet[][for NGC\,1425]{Mould:etal:00}.}
\end{deluxetable}

\clearpage
\begin{deluxetable}{lcccrccccc}
\tablewidth{0pt}
\tabletypesize{\footnotesize}
\tablecaption{Parameters of the Three Nearest Galaxy Groups.\label{tab:08}}   
\tablehead{
 \colhead{Group} & 
 \colhead{$N$} & 
 \colhead{$\langle D\rangle$} & 
 \colhead{$\langle D_{0}\rangle$} & 
 \colhead{$\langle v_{\odot}\rangle$} & 
 \colhead{$\langle v_{0}\rangle$} & 
 \colhead{$\langle v_{00}\rangle$} & 
 \colhead{$\langle v_{220}\rangle$} & 
 \colhead{$H_{i}$} & 
 \colhead{$\Delta v_{\rm Hubble}$} \\
 \colhead{(1)} & 
 \colhead{(2)} & 
 \colhead{(3)} & 
 \colhead{(4)} & 
 \colhead{(5)} & 
 \colhead{(6)} & 
 \colhead{(7)} & 
 \colhead{(8)} & 
 \colhead{(9)} & 
 \colhead{(10)}  
} 
\startdata
M\,81   & 29 & 3.69 & 3.50 &  51$\pm$20 & 211 & 213 & 234 & 66.8$\pm$6 & 23 \\ 
Cen\,A  & 20 & 3.75 & 4.29 & 536$\pm$26 & 261 & 262 & 288 & 67.1$\pm$6 & 29 \\ 
IC\,342 &  7 & 3.36 & 2.94 &  12$\pm$18 & 256 & 258 & 290 & 98.5$\pm$6 & 113 \\
\enddata
\end{deluxetable}

\end{document}